\documentclass[preprint,5p,times,twocolumn,nopreprintline]{elsarticle}

\usepackage{amssymb}
\usepackage{amsmath}
\usepackage{graphicx}
\usepackage{adjustbox} 
\usepackage{booktabs}
\usepackage{multirow}
\usepackage{xcolor}

\usepackage{siunitx}
\usepackage{array}
\usepackage{tabularx}
\usepackage{enumitem}
\usepackage{comment}
\usepackage{tabularx}
\usepackage{multirow}
\usepackage{lineno}
\usepackage{setspace}

\newcolumntype{L}{>{\raggedright\arraybackslash}X}
\newcommand{\tablesetup}{\small
  \setlength{\tabcolsep}{4pt}\renewcommand{\arraystretch}{1.1}}
\newcommand{\tablesetuptight}{\footnotesize
  \setlength{\tabcolsep}{3pt}\renewcommand{\arraystretch}{1.15}}
\usepackage{url}

\usepackage[
  hidelinks,
  breaklinks=true,
  bookmarksnumbered=true,
  pdftitle={Development of a real-time FPGA-based CNN trigger operating on full-granularity calorimeter images},
  pdfsubject={Nuclear Instruments and Methods in Physics Research Section A}
]{hyperref}

\newcommand{\kpiee}{\ensuremath{K^{0}_{L}\rightarrow\pi^{0}e^{+}e^{-}}}
\newcommand{\kpinn}{\ensuremath{K^{0}_{L}\rightarrow\pi^{0}\nu\bar{\nu}}}
\newcommand{\kpipipi}{\ensuremath{K^{0}_{L}\rightarrow\pi^{+}\pi^{-}\pi^{0}}}

\newcommand{\klong}{\ensuremath{K^{0}_{L}}}

\newcommand{\esig}{\ensuremath{\varepsilon_{\mathrm{sig}}}}
\newcommand{\Rbkg}{\ensuremath{R_{\mathrm{bkg}}}}

\hypersetup{pdftitle={Development of a real-time FPGA-based CNN trigger operating on full-granularity calorimeter images}}

\journal{Nuclear Instruments and Methods in Physics Research Section A}

\makeatletter
\ifpreprint
  \let\originalpprintMaketitle\pprintMaketitle
  \renewcommand{\pprintMaketitle}{%
    \edef\frontmatterholdinginserts{\the\holdinginserts}%
    \ifLineNumbers\global\holdinginserts=3\relax\fi
    \originalpprintMaketitle
    \global\holdinginserts=\frontmatterholdinginserts\relax}
  \myfooter[L]{\parbox[t]{\textwidth}{%
    Preprint submitted to \@journal\endgraf
    \hfill\@date}}
\fi
\makeatother

\begin{document}

\begin{frontmatter}

%==============================================================================
% TITLE (NIM A sentence case; KOTO~II with non-breaking space)
%==============================================================================
\title{Development of a real-time FPGA-based CNN trigger \\ operating on full-granularity calorimeter images}

\author[nknu]{J.-J.~Wu}
\author[kek]{Y.-T.~Lai}
\author[nknu]{Y.-C.~Tung\corref{cor1}}
\ead{yctung@mail.nknu.edu.tw}
\cortext[cor1]{Corresponding author}
\author[nknu]{S.-T.~Chen}
\author[nknu]{Y.-S.~Liu}

\affiliation[nknu]{
    organization={Department of Physics, National Kaohsiung Normal University},
    city={Kaohsiung},
    postcode={824},
    country={Taiwan}
}

\affiliation[kek]{
    organization={Institute of Particle and Nuclear Studies, High Energy Accelerator Research Organization (KEK)},
    city={Tsukuba, Ibaraki},
    postcode={305-0801},
    country={Japan}
}

%==============================================================================
% ABSTRACT
%==============================================================================
\begin{abstract}

We present a real-time field-programmable gate array (FPGA) implementation of a convolutional neural network (CNN) for event classification on full calorimeter images that retain crystal-level shower information. To meet the resource and latency constraints of trigger electronics, the network parameters and hidden activations were quantized to 8 bits with quantization-aware training, and a lossless space-to-depth reorganization rearranged the input image into fewer spatial positions, reducing the estimated CNN inference latency by more than an order of magnitude. The approach was evaluated with the J-PARC KOTO detector, recorded data, and data-acquisition (DAQ) constraints as a benchmark, targeting the online separation of the rare \kpiee{} decay from its dominant \kpipipi{} background. As a proof of principle, the firmware was implemented and validated on an AMD Versal Premium VPK120 evaluation board. On 100{,}000 test events, inference on the board rejected $(99.34\pm0.04)\%$ of the recorded background sample enriched in \kpipipi{} decays at a nominal \kpiee{} efficiency of 95\%,
with a measured CNN inference latency of 17.9~\si{\micro\second}. Combining the measured image-assembly and CNN inference latencies with the estimated link-transfer and score-return times gives a projected end-to-end latency of \SI{25.9}{\micro\second}, within the adopted \SI{32}{\micro\second} DAQ latency budget. These results demonstrate the feasibility of using full-granularity calorimeter images for CNN-based event selection that achieves high background rejection while keeping the trigger latency within tens of microseconds.

\end{abstract}


\end{frontmatter}

%==============================================================================
\section{Introduction}
\label{sec:intro}

%==============================================================================

Convolutional neural networks (CNNs) are widely used for offline data
analysis in high-energy physics~\cite{radovic2018}.
In detectors with spatially segmented readout, particle interactions
produce energy-deposition patterns that can be represented as images.
CNNs can learn spatial features from these images for event classification
and particle identification, such as separating electrons from charged
pions~\cite{belayneh2020} and distinguishing neutron-induced from
photon-induced calorimeter clusters~\cite{csd}.
These capabilities are also highly desirable in online trigger systems,
where effective background rejection is needed to reduce the
data-acquisition (DAQ) load while retaining events of physics interest.
Applying CNNs at the trigger stage, however, requires inference to be
completed within the trigger latency budget.
Implementations on field-programmable gate arrays (FPGAs) must also fit
within the available logic, memory, and arithmetic resources.

Advances in high-level synthesis (HLS) and quantization-aware training, together with tools such as hls4ml, have enabled low-latency neural-network inference on FPGAs~\cite{hls4ml,hls4ml_cnn,qkeras}.
For calorimeters, FPGA implementations include electron--pion identification and energy regression from hit clusters in a simulated calorimeter~\cite{garnet}, and clustering and per-cluster classification from aggregated trigger cells at Belle~II~\cite{belle2_calo_gnn}.
Brooke et al.~\cite{brooke_cnn} classified collision events from calorimeter-tower images for a topological trigger and selected a model with downsampled inputs.
Here, we focus on event classification using full calorimeter images that retain shower information at the granularity of individual crystals.
The question addressed here is how to achieve high background rejection at a fixed signal efficiency within a trigger latency budget of tens of microseconds, and whether the performance of the quantized software model is maintained in hardware.

This study is motivated by the search for the rare decay \kpiee{}
% REF-AUDIT 2026-10-16: added tung2020 to support the KOTO pi0ee search motivation.
in the KOTO experiment~\cite{koto_pnn,tung2020} at J-PARC and its planned
successor, KOTO~II~\cite{koto2,tung2024}.
The decay is a sensitive probe of charge-parity ($CP$)
violation~\cite{pi0ee_sm1,pi0ee_sm2} and complements the primary
\kpinn{} program of both experiments.
The Standard Model branching ratio is of order $10^{-11}$~\cite{pi0ee_sm2}, about an order of magnitude below the experimental upper limit set by the Fermilab KTeV experiment~\cite{ktev}.
At the event rates needed to reach this sensitivity, a major challenge for the trigger is rejecting the abundant \kpipipi{} background, whose branching ratio is $12.5\%$~\cite{pdg}.
After $\pi^{0}\rightarrow\gamma\gamma$, both signal and background
contain two photons and two charged particles and can satisfy trigger
conditions based on energy and particle multiplicity.
Electrons and positrons, however, typically produce compact electromagnetic 
showers, whereas charged pions that interact
hadronically produce broader and more fragmented energy deposits, although they
can also leave minimum-ionizing or electromagnetic-like signatures. 
These differences motivate event classification using calorimeter
images without additional particle-identification detectors.

KOTO provides the experimental benchmark for this study.
Its 2,716-channel CsI calorimeter~\cite{csi_calorimeter} defines the
input image, while its readout and trigger architecture~\cite{cluster_trigger}
sets the DAQ constraints (Sec.~\ref{sec:benchmark}).
The CNN uses calorimeter shower patterns to distinguish \kpiee{}
signal events from \kpipipi{} background events
(Sec.~\ref{sec:cnn}).
Using this benchmark, we developed proof-of-principle FPGA firmware
for full-image event classification and validated it on a commercial
FPGA evaluation board, hereafter referred to as the Deep Learning
Module (DLM).
To meet the latency and resource constraints, the CNN was quantized using quantization-aware
training~\cite{qkeras}, and lossless
space-to-depth reorganization~\cite{yolo9000,spdconv} was applied to shorten
inference latency.
We quantify the resulting trade-offs in Sec.~\ref{sec:opt} and present
the FPGA implementation and hardware validation in Sec.~\ref{sec:impl}.
Finally, we estimate the reduction in the trigger rate for the KOTO
\kpiee{} trigger line and assess hardware scalability toward the larger
KOTO~II calorimeter (Sec.~\ref{sec:disc}).

%==============================================================================
\section{Detector, trigger, and design requirements}
\label{sec:benchmark}
%==============================================================================

\subsection{Calorimeter and trigger}
\label{sec:benchmark:detector}

\begin{figure}[tbp]
    \centering
    \includegraphics[width=\linewidth]{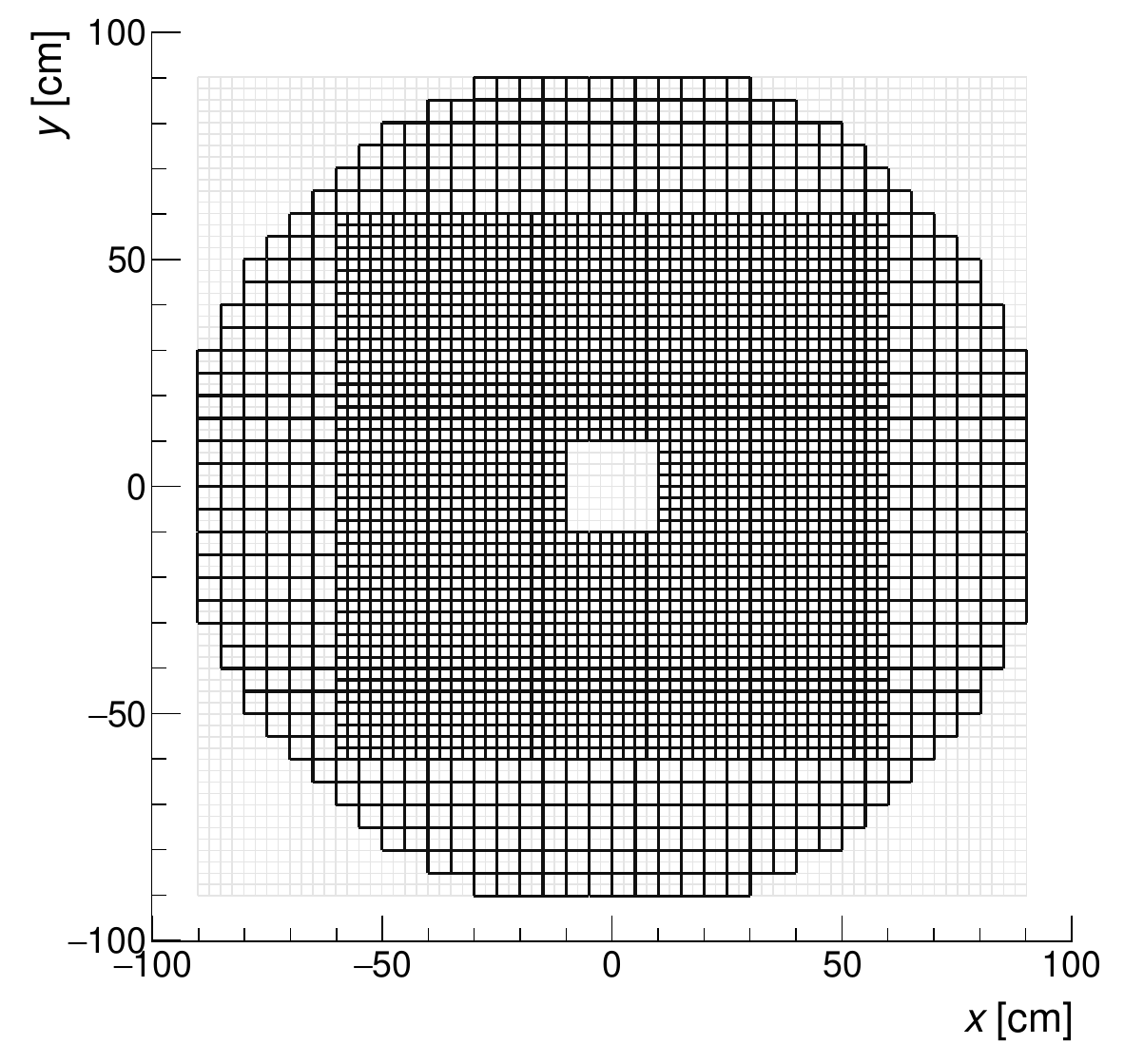}
    \caption{
	Transverse layout of the KOTO CsI calorimeter. The central region
    contains 2,240 $25\times25$~mm$^2$ crystals and is surrounded by 476
    $50\times50$~mm$^2$ crystals. The central square aperture is the neutral-beam
    hole. The light-gray $25\times25$~mm$^2$ grid shows the $72\times72$ cells
    used to construct the CNN input.
    }
    \label{fig:KOTODetector}
\end{figure}

The KOTO electromagnetic calorimeter consists of 2,716 undoped CsI crystals,
each read out independently to provide fine-grained measurements of
shower development~\cite{csi_calorimeter,koto_detector}.
As shown in Fig.~\ref{fig:KOTODetector}, the central region contains 2,240
crystals with a cross section of $25\times25$~mm$^2$, surrounded by 476
crystals with a cross section of $50\times50$~mm$^2$. A square hole at the
center allows the neutral beam to pass through. 
The CsI signals are digitized at \SI{125}{MHz}. 
Channel energies for the training, validation, and test samples were obtained from the standard offline reconstruction, which integrates pedestal-subtracted waveforms over a 64-sample window defined relative to the pulse time and applies timing and gain calibrations.
Signals from the other detector subsystems enter the upstream trigger
conditions but are not inputs to the DLM.

%==============================================================================

\begin{figure}[tbp]
  \centering
\includegraphics[width=.333333\linewidth]{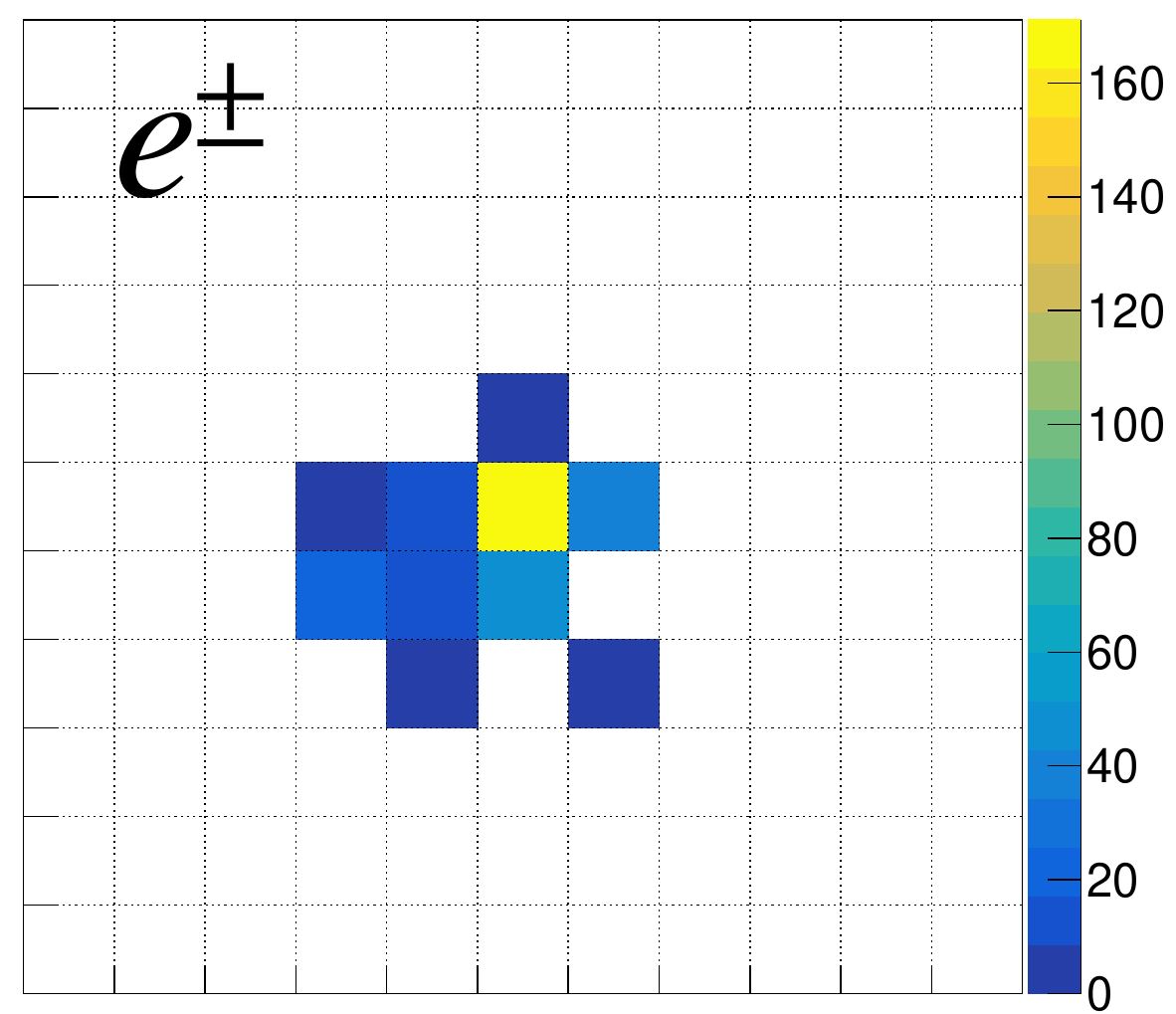}%
\includegraphics[width=.333333\linewidth]{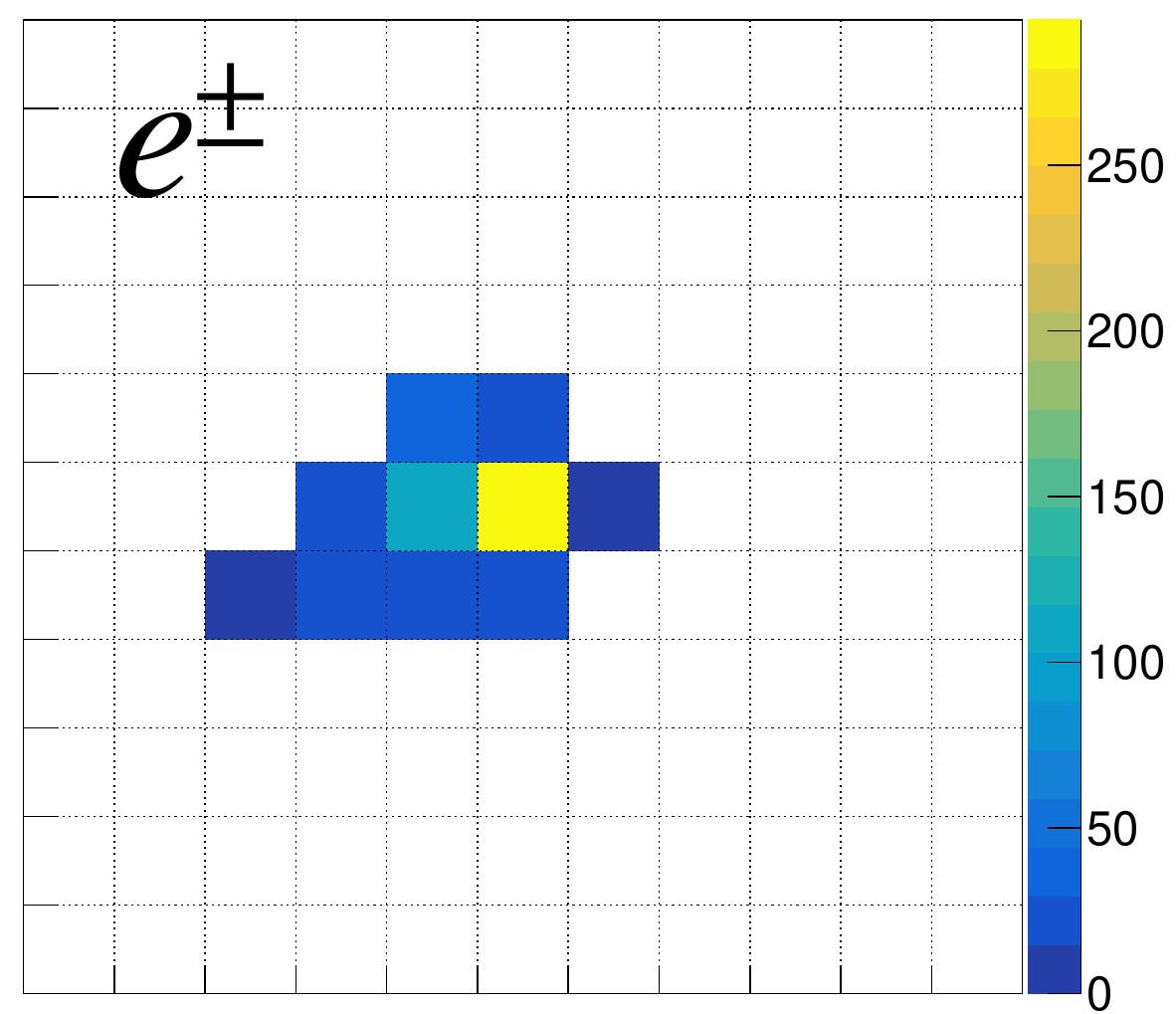}%
\includegraphics[width=.333333\linewidth]{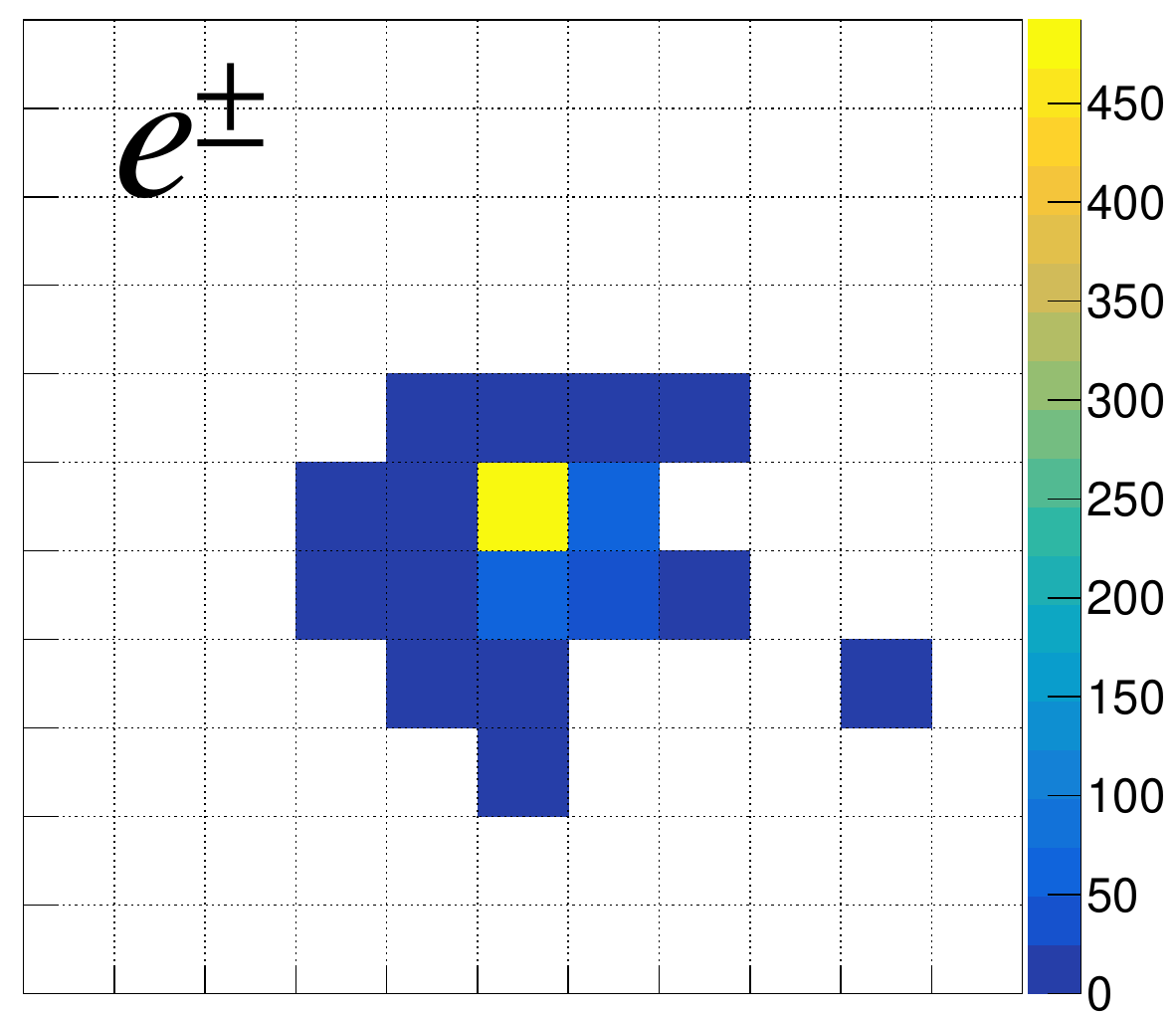}

\vspace{1mm}

\includegraphics[width=.333333\linewidth]{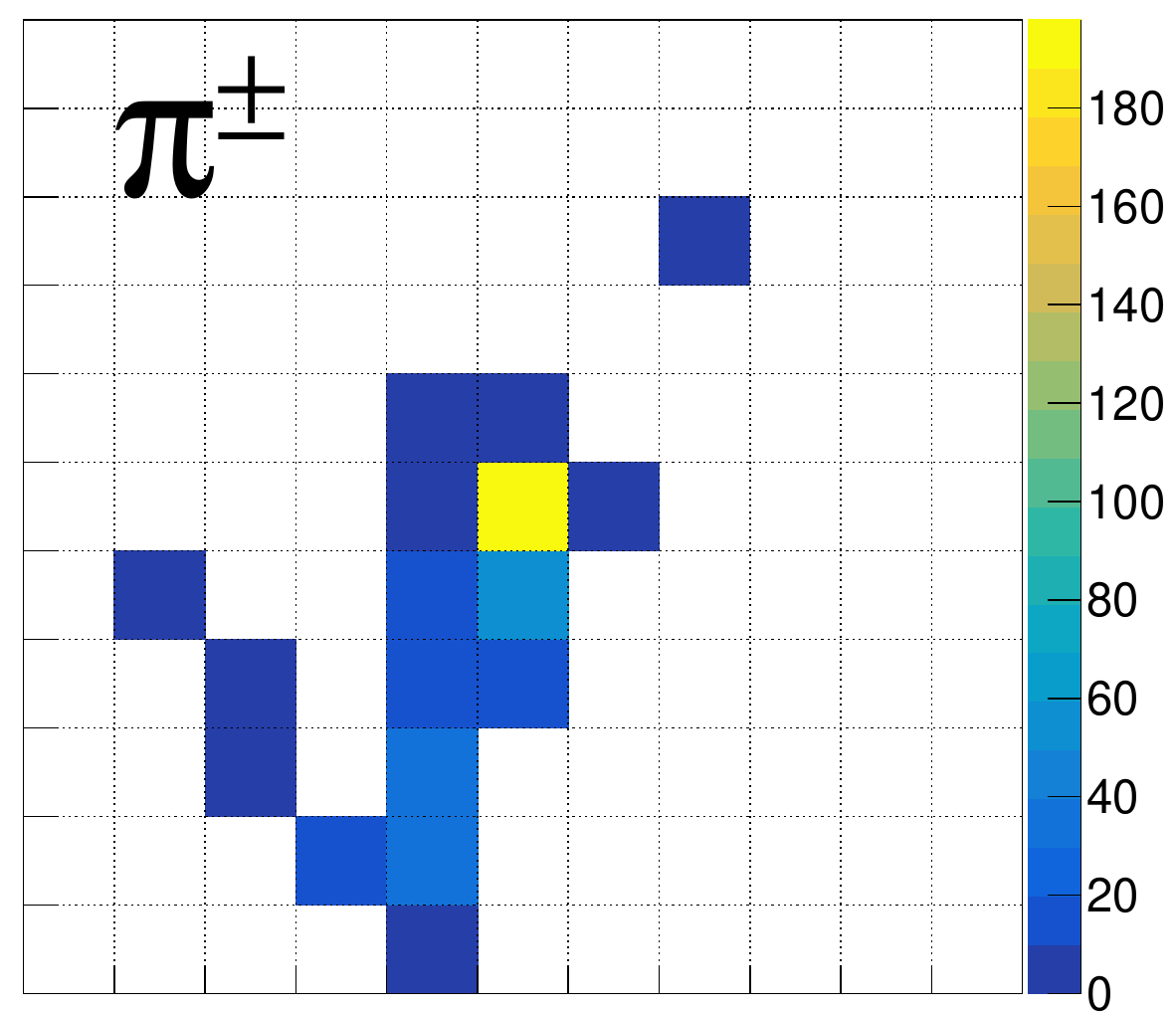}%
\includegraphics[width=.333333\linewidth]{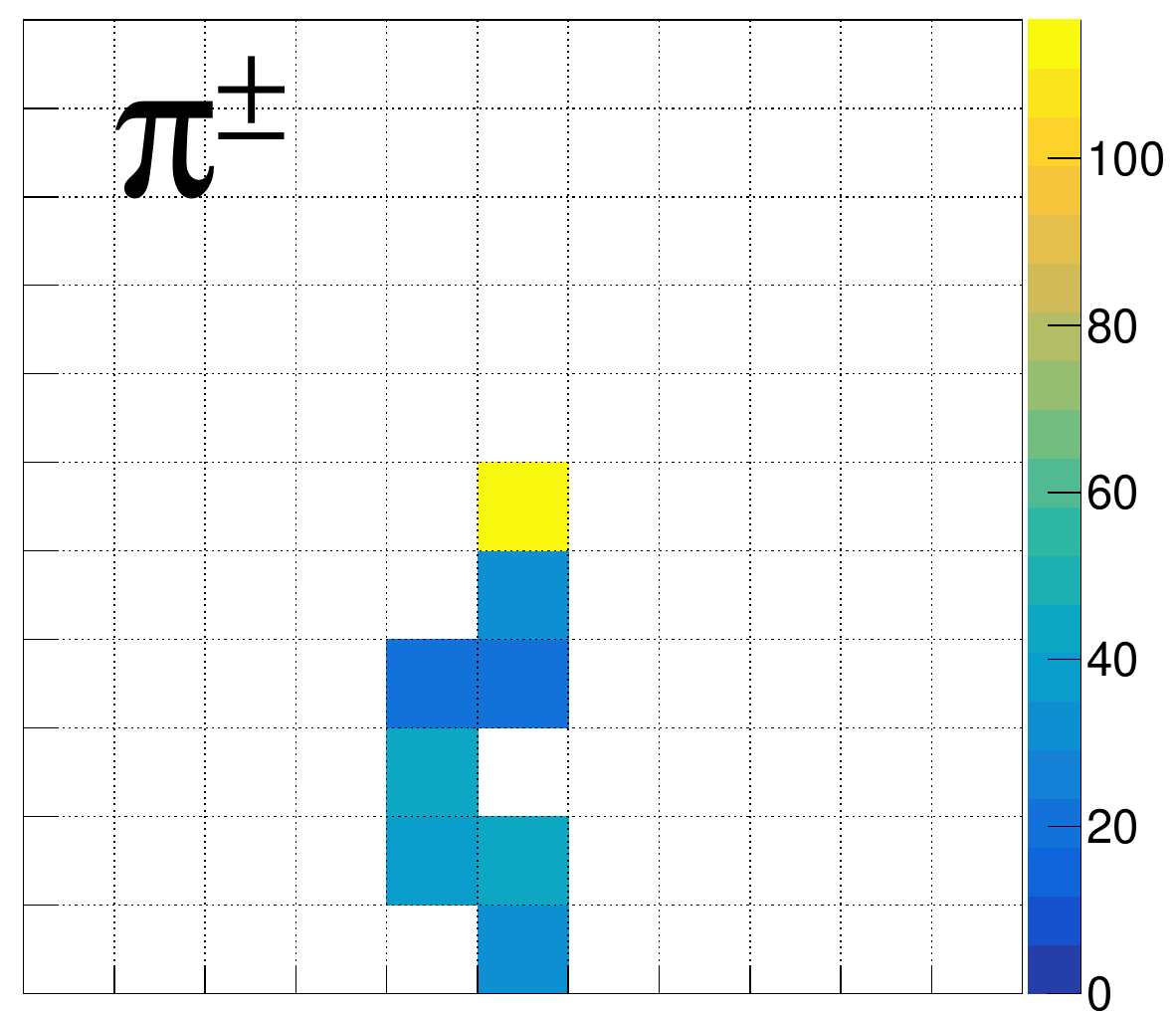}%
\includegraphics[width=.333333\linewidth]{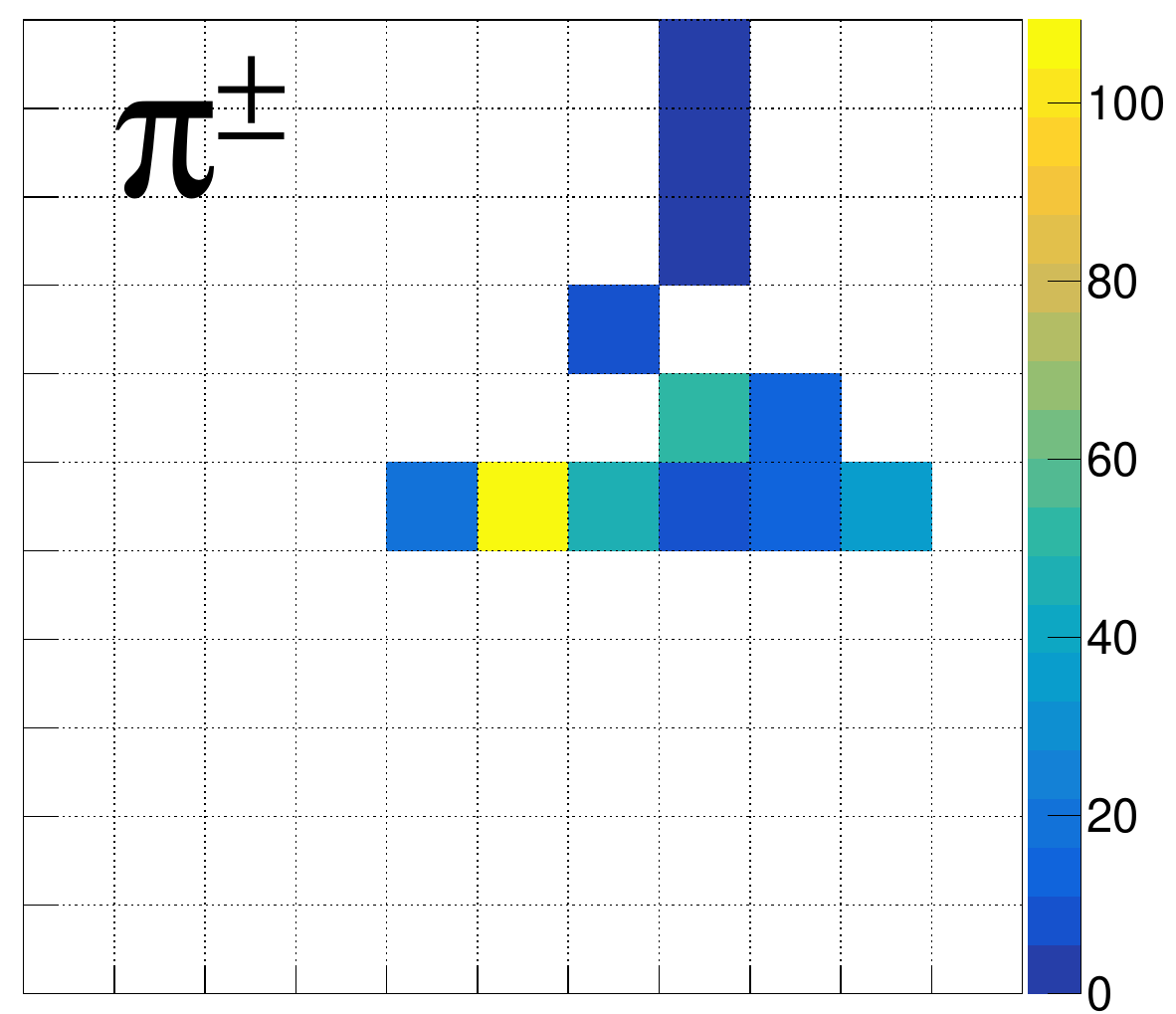}

\caption{Examples of electron-induced clusters from simulated \kpiee{} events (top) and
charged-pion-induced clusters from recorded background events (bottom).
 Each panel shows a $12\times12$ region of $25\times25$~mm$^2$ cells
around one cluster.
The color scale gives the cell energy in MeV and is set independently for each
panel.}
  \label{fig:images}
\end{figure}

Two hardware trigger levels, Level-I and Level-II, define the event stream
that the DLM would process. The Level-I decision requires a total energy above
550~MeV in the CsI calorimeter, together with conditions on the surrounding
photon and charged-particle counters.
For the \kpiee{} trigger line, an energy deposit is required in the
charged-veto counter in front of the calorimeter to retain events containing
the signal $e^+e^-$ pair.
Events with in-time signals in the other veto counters are rejected.
Level-II applies a cluster-counting algorithm~\cite{cluster_trigger} to the
calorimeter energy deposits and selects events with four clusters. Both \kpiee{}
and \kpipipi{} can satisfy this condition, so an additional selection using
the calorimeter energy distribution is needed to distinguish them.
Figure~\ref{fig:images} shows representative electron and charged-pion clusters
from the signal and background samples, illustrating the shower shapes
available for discrimination.

%==============================================================================

\subsection{System-level design requirements}
\label{sec:benchmark:req}

For each event accepted by Level-I and Level-II, the DLM would
receive the energies of all 2,716 CsI channels and return the corresponding CNN score.
The design was evaluated against three requirements.

\begin{figure}[tbp]
  \centering
  \includegraphics[width=\linewidth]{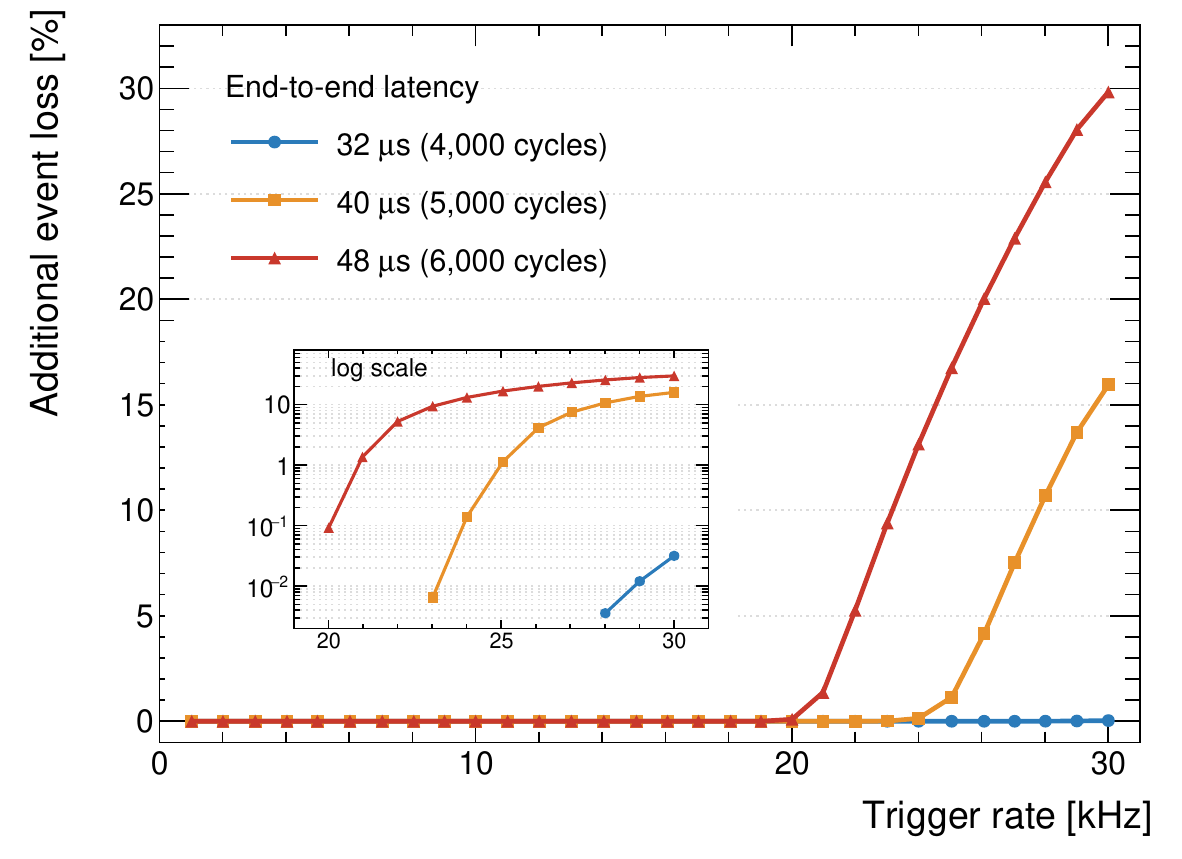}
  \caption{Difference in event-loss fraction, expressed in percentage points, 
  between the DAQ with and without the
  DLM, evaluated for different end-to-end module latencies using the DAQ
  emulator. Trigger rate denotes the mean rate during the \SI{2}{s} beam spill. Cycle counts refer to the \SI{125}{MHz} DAQ clock. The
  inset repeats the same quantity on a logarithmic vertical scale. Latencies
  below  \SI{32}{\micro\second} ($4{,}000$ cycles) gave no additional loss 
  within the emulator precision and are omitted.}
  \label{fig:deadtime}
\end{figure}

\begin{itemize}[leftmargin=*,itemsep=2pt,topsep=2pt]
\item \textbf{Latency requirement (R1):}
As a third-level trigger, the DLM must return the CNN score before the event is forwarded from the intermediate buffer of the DAQ
data-transfer pipeline to the next stage. 
In the KOTO DAQ, forwarding an event takes about
8,000 cycles of the \SI{125}{MHz} system clock
(\SI{64}{\micro\second}). Two output ports handle successive events in
alternation, giving an effective interval of approximately
\SI{32}{\micro\second} per event.
This interval was adopted as the end-to-end latency budget. It must accommodate
the transfer of the channel energies to the DLM, image assembly and
normalization, CNN inference, and return of the score. 

The effect of this latency on event loss was evaluated with a DAQ emulator,
in which trigger requests follow a Poisson process within each
beam spill, the slow-extraction period of about \SI{2}{s} that repeats every
\SI{4.2}{s}.
The emulator uses a non-overlapping service model, with the service time
set to the DLM latency.
Figure~\ref{fig:deadtime} shows the additional event loss relative to the
same DAQ without the  DLM. This loss remained negligible for the simulated
latencies up to \SI{32}{\micro\second}
at trigger rates up to  \SI{30}{kHz}, but became significant at  40 and
\SI{48}{\micro\second}. These results support the adopted budget under the
assumptions of the emulator. 

\item \textbf{Classification performance (R2):}
 A working point with signal efficiency $\esig=95\%$ was adopted throughout this study, and the
background rejection \Rbkg{} at this efficiency is the primary
classification metric. The design target is $\Rbkg\geq90\%$, corresponding
to at least tenfold background suppression. 

\item \textbf{Resources (R3):}
The complete DLM implementation, including image assembly, CNN inference,
score output, and communication interfaces, must fit within a single FPGA, with sufficient capacity reserved for the surrounding trigger infrastructure and future system integration. 
The demonstration used the AMD Versal Premium XCVP1202 device on the
VPK120 evaluation board~\cite{vpk120}.

\end{itemize}

%==============================================================================
\section{CNN-based event classification}
\label{sec:cnn}
%==============================================================================

\subsection{Dataset and image representation}
\label{sec:cnn:image}

The signal decay \kpiee{} was simulated 
with the matrix element of Ref.~\cite{greenlee} using the KOTO 
\textsc{Geant4}~\cite{geant4} framework. Neutral kaons were generated according
to the measured momentum spectrum~\cite{kl_spectrum}, decayed within the
fiducial volume, and their decay products were propagated through the detector-response and electronics simulation.
To reproduce the detector environment during physics data taking,
accidental activity was included by overlaying recorded random-trigger waveforms
onto the simulated events~\cite{koto_pnn}.

The background sample was taken from recorded KOTO events enriched in \kpipipi{} decays. 
Hadronic shower development is more difficult to model
reliably than electromagnetic shower development, so this choice provides
pion-shower patterns measured under the experimental conditions.
The quoted rejection is conditional on this selected background sample.
Because signal and background originate from different sources, the classifier
could also exploit residual differences between simulation and data.
The control-sample study in Sec.~\ref{sec:cnn:datamc} examines
this effect through a comparison of the electromagnetic response.

Signal and background were subjected to a common offline selection.
The signal simulation includes the Level-II cluster-counting response.
The same offline veto requirements were applied to both samples, with
thresholds tighter than the corresponding Level-I settings. 
Approximately $1.1\times10^{6}$ events passing this selection were used
in this study. Of these,
100{,}000 were reserved as a test sample, and the remainder was
divided into training and validation sets of approximately 800{,}000 and
200{,}000 events, respectively. Each subset contains equal numbers of signal
and background events.
For the DLM models, the validation set was used to choose hyperparameters,
per-layer numerical precisions, and training checkpoints. The common test sample was
used to compare classification performance and to determine score thresholds
at fixed signal efficiencies.

The CNN input was constructed from the per-channel CsI energies
introduced in Sec.~\ref{sec:benchmark:detector}.
These energies were arranged on
a uniform $72\times72$ grid at the $25\times25$~mm$^2$ pitch of the central
crystals, shown as the fine gray grid in the background of
Fig.~\ref{fig:KOTODetector}. Each
$50\times50$~mm$^2$ outer crystal was represented by four cells, with its energy
distributed equally among them. Cells corresponding to the beam hole or regions
outside the calorimeter were assigned zero.
The resulting single-channel tensor has dimensions $72\times72\times1$.
Each cell energy was then divided by the largest cell energy in the event
after outer-crystal subdivision, which maps the input to $[0,1]$ and sets
the most energetic cell to unity. 
Because the same factor is applied to all cells, this
normalization removes the overall energy scale while preserving the spatial
pattern, relative cell energies, and cluster multiplicity.

\subsection{CNN model and floating-point performance}
\label{sec:cnn:model}

The floating-point baseline was implemented as a CNN using Keras with
TensorFlow version 2.14.1~\cite{tensorflow}. It consists of three
convolutional blocks with 32, 64, and 128 filters, respectively. Each block
contains an unpadded $2\times2$ convolution with unit stride, a rectified
linear unit (ReLU) activation, and $2\times2$ max pooling with stride two.
The resulting $8\times8\times128$ feature map is flattened and passed to a
dense layer with 128 neurons and a ReLU activation. A final sigmoid output provides the signal
score $y\in[0,1]$.
The network contains 1{,}090{,}145 trainable parameters, most of which belong
to the hidden dense layer. 
Training used binary cross-entropy and the Adam optimizer~\cite{adam}
with a batch size of 512, an $L_2$ penalty of $5\times10^{-5}$ on the
convolutional and dense kernels, and dropout with a rate of 0.05 after the
hidden dense layer. 
 The learning rate was scheduled to increase linearly to $5\times10^{-4}$
over the first five epochs and then decrease toward $10^{-6}$ at
epoch 180, the maximum number of training epochs.
Training used an early-stopping patience of 30 epochs.
The retained checkpoint was the one with the highest validation-set
background rejection at $\esig=95\%$.
No class weighting or data augmentation  was applied.

Classification performance is characterized by the fraction of background
events rejected at a fixed signal efficiency. For a threshold $t$, an event is
accepted if its output score satisfies $y\geq t$. The signal efficiency
and background rejection are then defined as
$\esig(t)=N_{\mathrm{sig}}(y\geq t)/N_{\mathrm{sig}}$ and
$\Rbkg(t)=1-N_{\mathrm{bkg}}(y\geq t)/N_{\mathrm{bkg}}$, where the
denominators are the numbers of labeled signal and background events in the
sample being evaluated.
 For each model, the threshold was scanned on the test sample to evaluate
rejection at the nominal $\esig=95\%$ working point and at 98\% and 90\%
for comparison.

\begin{figure}[tbp]
  \centering
  \includegraphics[width=1.0\linewidth]{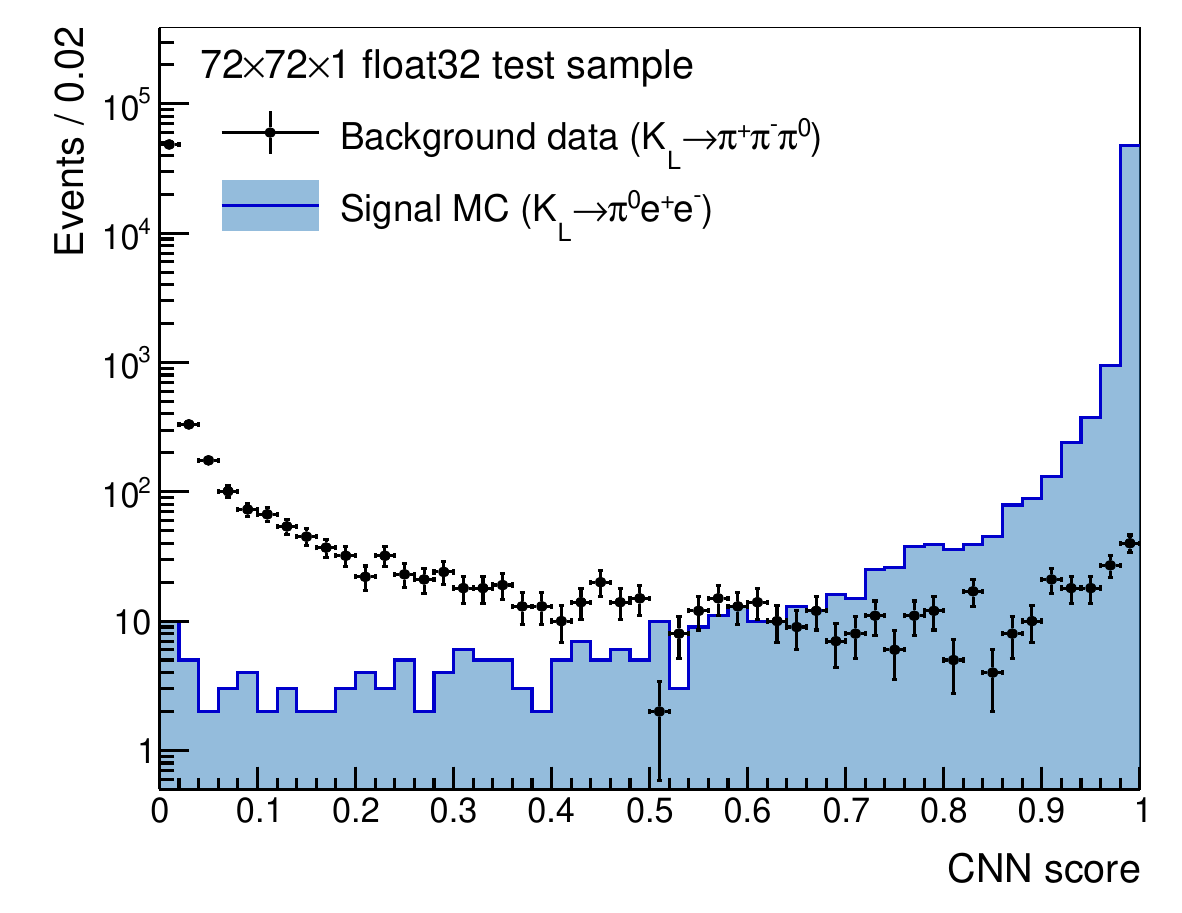}
  \caption{%
  Score distributions for 50{,}000 \kpiee{} signal events from Monte Carlo
  (MC) simulation and 50{,}000 recorded background events enriched in
  \kpipipi{} decays, evaluated with the float32 $72\times72\times1$ baseline
  model on the independent test sample.}
  \label{fig:score}
\end{figure}

Figure~\ref{fig:score} shows the score distributions for the floating-point
baseline. 
For the $72\times72\times1$ input, the baseline gave
$\Rbkg=(99.83\pm0.02)\%$, $(99.92\pm0.01)\%$, and
$(99.96\pm0.01)\%$ at $\esig=98\%$, 95\%, and 90\%, respectively.
Unless stated otherwise, the quoted uncertainties are binomial
standard errors evaluated using the 50{,}000 background test events
at the specified threshold.
The floating-point baseline met the R2 rejection target on the
labeled test sample.

\subsection{Control-sample comparison of data and simulation}
\label{sec:cnn:datamc}

Because the \kpiee{} signal sample was simulated while the \kpipipi{} background 
was taken from data, 
an independent four-cluster control sample was used to 
examine how well the network response is reproduced by the simulation. 
The control sample consists mainly of
$\klong\rightarrow 2\pi^{0}$ decays collected with the
same Level-I and Level-II conditions as the \kpiee{} line, except that the
charged-veto counter in front of the calorimeter rejects rather than selects
events with a charged particle.
The standard offline selections~\cite{koto_pnn} were applied to suppress 
$\klong\rightarrow3\pi^{0}$ contamination present in the data.
The four photons in $\klong\rightarrow2\pi^{0}$ decays 
initiate electromagnetic showers in the
CsI calorimeter, so the comparison probes the simulated photon-induced
response seen by the classifier.
For each of the $\klong\rightarrow2\pi^{0}$ and
$\klong\rightarrow3\pi^{0}$ decay modes, $4\times10^{6}$ events were
simulated.
 After all selections, 195{,}039 $\klong\rightarrow2\pi^{0}$ and
4 $\klong\rightarrow3\pi^{0}$ simulated events remained.
The selected data sample contained 14{,}295 events.
The same input construction and network inference
were applied to both the data and simulated samples. 
The simulated $\klong\rightarrow2\pi^{0}$ and $\klong\rightarrow3\pi^{0}$ 
modes were subsequently normalized by their
branching ratios~\cite{pdg} ($8.64\times10^{-4}$ and $0.1952$, respectively), and 
one common scale factor then normalized the total simulated
distribution to the observed data yield. 

\begin{figure}[tbp]
  \centering
  \includegraphics[width=1.\linewidth]{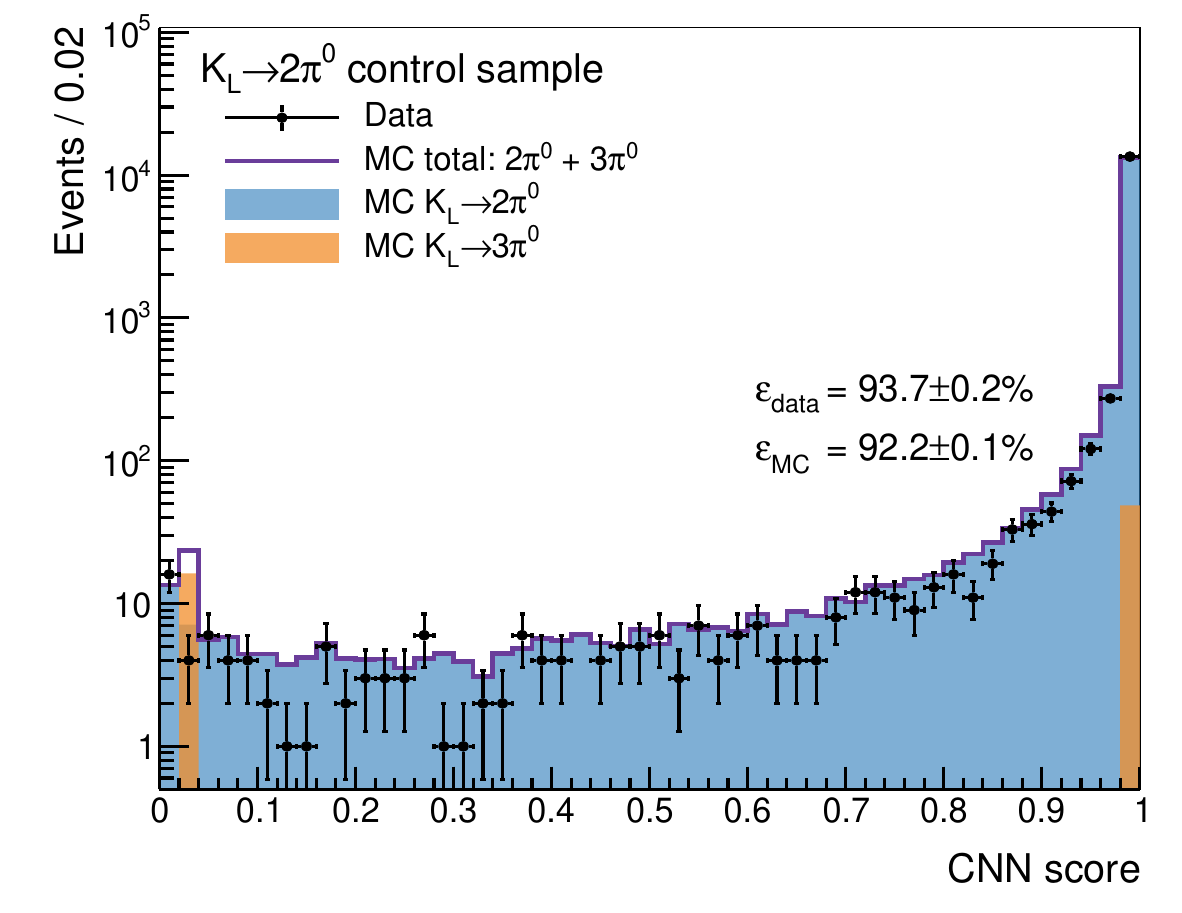}
	\caption{%
	CNN-score distributions for the
	$\klong\rightarrow 2\pi^{0}$ control sample in data
	(black points) and the MC simulations.
	The filled histograms show the
	$\klong\rightarrow 2\pi^{0}$ (blue) and
	$\klong\rightarrow3\pi^{0}$ (orange) components.
	The solid purple curve shows their sum, 
	whose total area was normalized to the data.
	The scores were evaluated with the float32
	$72\times72\times1$ model.
	The quoted efficiencies were evaluated at the score threshold
	corresponding to 95\% efficiency for the \kpiee{} signal test sample.}
  \label{fig:k2pi0}
\end{figure}

Figure~\ref{fig:k2pi0} compares the float32 baseline scores for 
data and simulated control events. 
Applying the baseline threshold
 corresponding to 95\% efficiency for the
\kpiee{} signal gave control-sample selection efficiencies of
$(93.71\pm0.20)\%$ in data and $(92.17\pm0.09)\%$ in the 
simulation, where 
the quoted uncertainties are statistical errors. 
The simulation reproduced the overall shape of the control-sample
score distribution, with a selection efficiency 1.54 percentage points
lower than in data.
The control and signal final states differ in relative cluster energies
and event topologies, so the control-sample efficiencies need not equal
95\%.
This comparison provides indirect support for the use of simulated
signal events for training and efficiency evaluation.

%==============================================================================
\section{FPGA-oriented optimization}
\label{sec:opt}
%==============================================================================

\subsection{Quantization}
\label{sec:opt:quant}

Reduced-precision fixed-point representations require less parameter
storage and fewer arithmetic resources than floating-point representations.
The CNN was quantized using
quantization-aware training (QAT) implemented with QKeras~\cite{qkeras}
(version 0.9.0). During training, the forward pass emulates the reduced
precision of the weights and activations so that the model can adapt to
quantization. 
Fixed-point precision is denoted $\langle W,I\rangle$ following the
\texttt{ap\_fixed} convention used by hls4ml: $W$ is the total word length and
$I$ includes the sign bit for signed quantities, leaving $W-I$ fractional
bits and a resolution of $2^{-(W-I)}$~\cite{vitis}.
Weights, biases,
and accumulators are signed. ReLU outputs are non-negative and
use the corresponding unsigned fixed-point type, written
$\mathrm{u}\langle W,I\rangle$, in which all $I$ integer bits carry magnitude.    
A model described as $n$-bit uses $n$-bit trainable parameters and hidden
activations, while its input and score interfaces remain
$\langle16,2\rangle$. Since the per-event normalization of
Sec.~\ref{sec:cnn:image} maps every cell into $[0,1]$, this input format
resolves increments of $2^{-14}$ in normalized cell energy. 
Keeping the interfaces fixed allows the
parameter and activation precisions to be compared without changing
the input representation.
Table~\ref{tab:quant} uses a common integer width $I$ 
for the parameters and hidden activations of
each model to establish a global precision baseline. 
Each model was retrained separately, so differences
among the fixed-point results also include training variation.

Table~\ref{tab:quant} compares the QKeras test-set results for the
$72\times72\times1$ network at fixed-point precisions from 16 down to 2 bits
with the float32 baseline. The global $\langle8,2\rangle$ model reached
$\Rbkg=99.89\%$ at $\esig=95\%$, compared with 99.92\% for both the 16-bit model
and the floating-point baseline.
The 8-bit model requires only one-quarter of 
the parameter storage of the float32 baseline.
Rejection decreased substantially at four bits and fell sharply at two bits. 
The global $\langle8,2\rangle$ format was therefore adopted for the space-to-depth
comparison.

\begin{table}[tbp]
\centering
\caption{Background rejection at nominal signal efficiencies for the
$72\times72\times1$ input, comparing the float32 baseline with the
quantized models.
The parameter bits count the trainable weights and biases only.
The precision $\langle W,I\rangle$ is defined in the text.
The binomial standard errors in $\Rbkg$ were at most 0.021 percentage
points for the float32, 16-bit, and 8-bit rows.}
\label{tab:quant}
\tablesetup
\begin{adjustbox}{max width=\linewidth}
\begin{tabular}{@{}lccccc@{}}
\toprule
& & & \multicolumn{3}{c}{$\Rbkg$ at $\esig$} \\
\cmidrule(lr){4-6}
Model & Precision & Parameter bits & 98\% & 95\% & 90\% \\
\midrule

Keras
& float32
& 34{,}884{,}640
& 99.83\%
& 99.92\%
& 99.96\% \\

QKeras
& $\langle16,6\rangle$
& 17{,}442{,}320
& 99.82\%
& 99.92\%
& 99.96\% \\

QKeras
& $\langle8,2\rangle$
& 8{,}721{,}160
& 99.78\%
& 99.89\%
& 99.94\% \\

QKeras
& $\langle4,2\rangle$
& 4{,}360{,}580
& 94.82\%
& 96.98\%
& 98.16\% \\

QKeras
& $\langle2,1\rangle$
& 2{,}180{,}290
& 11.64\%
& 18.80\%
& 27.71\% \\

\bottomrule
\end{tabular}
\end{adjustbox}
\end{table}

\subsection{Space-to-depth reorganization}
\label{sec:opt:s2d}

\begin{figure*}[tbp]
  \centering
  \includegraphics[width=0.95\linewidth]{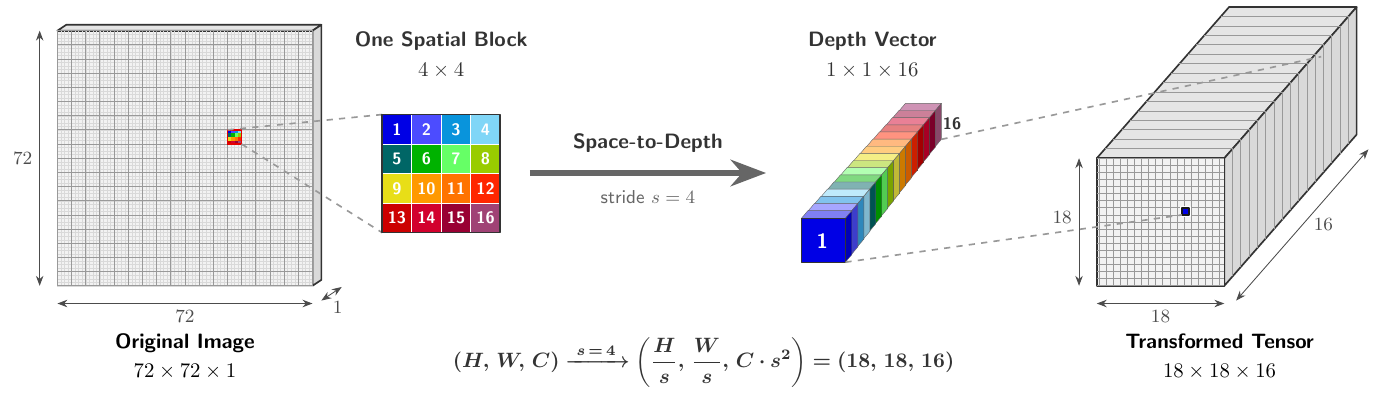}
  \caption{Space-to-depth transformation of the CsI image: each $s\times s$
  block of  cells is folded into $s^{2}$ input channels, shown for
  $s=4$, which maps the $72\times72\times1$ calorimeter input to an
  $18\times18\times16$ tensor without altering any cell value. $H$, $W$, and
  $C$ denote the tensor height, width, and channel count.}
  \label{fig:s2d}
\end{figure*}

In the selected streaming implementation, convolutional output positions
are processed sequentially. For the $72\times72\times1$ input, the first valid
$2\times2$ convolution evaluates $N_1=71\times71=5{,}041$ positions with
a single input channel. This combination of many output positions and
limited parallelism across input channels motivates changing the input
representation. The space-to-depth transformation reorganizes each
$s\times s$ spatial block into $s^2$ channels through an
information-preserving permutation used in computer
vision~\cite{yolo9000,spdconv}. This reduces the number of sequentially
processed positions while exposing parallelism across input channels
in the first convolution. Figure~\ref{fig:s2d} illustrates the $s=4$
case, which maps the $72\times72\times1$ image to an
$18\times18\times16$ tensor. The other evaluated strides and their
input dimensions are listed in Table~\ref{tab:s2dscan}.

For each input representation, the model was retrained using the same
stratified data split, convolutional widths of 32, 64, and 128, and a
hidden dense width of 128. The pooling layout was adjusted to accommodate
the smaller spatial dimensions, with all stages using $2\times2$ windows,
stride two, and no padding. Max pooling follows all three convolutions
(Conv0--Conv2)
for $s=1,2,3$, Conv0 and Conv1 for $s=4,6$, and Conv0 and Conv2 for $s=8$.
All models use signed $\langle8,2\rangle$ parameters, unsigned
$\mathrm{u}\langle8,2\rangle$ hidden activations, and signed
$\langle16,2\rangle$ input and score interfaces. The accumulators use
the uncompressed widths inferred by hls4ml for each model.
The reuse factor controls multiplier sharing, with larger values
reducing multiplier use at the cost of latency. Reuse factors were fixed
at 4, 16, 256, and 64 for Conv0, Conv1, the hidden dense layer, and the
output layer, respectively.
For Conv2, the reuse factor was 64 for $s=1,2,8$ and 128 for $s=3,4,6$.

Increasing $s$ reduces the number of first-layer spatial positions to
$N_1=(72/s-1)^2$. This shortens the first-layer latency, which dominates
the CNN-core latency in these implementations. Background rejection
also varied with $s$ because the corresponding models differ in spatial
connectivity, pooling layout, and network capacity. The resource
requirements did not decrease uniformly with latency because the $s^2$
input channels increase the multiplier and buffer costs.
Table~\ref{tab:s2dscan} summarizes the latencies obtained from the HLS
schedule, pre-implementation resource estimates, and background rejection
evaluated using AMD Vitis HLS~\cite{vitis} C simulation of all 100{,}000 test events.
In this common-format comparison, increasing $s$ from 1 to 4 reduced
the estimated core latency from  \SI{311.7}{\micro\second} to
 \SI{21.8}{\micro\second}, while background rejection at $\esig=95\%$
decreased from 99.90\% to 99.23\%. The $s=3$ core alone exceeds the
 \SI{32}{\micro\second} budget of R1, whereas $s=4$ leaves margin for
image assembly and external transfers, with estimated utilization below
25\% for every XCVP1202 resource type. The more aggressive $s=6$ and
$s=8$ foldings resulted in shorter latencies but lower background rejection.
The $s=4$ stride was selected for implementation based on
background rejection, latency, and resource constraints.

\begin{table*}[tbp]
\centering
\caption{
Vitis HLS C-synthesis estimates and C-simulation classification performance
for the space-to-depth representations at global $\langle8,2\rangle$
precision. The folding stride is $s$, and $N_1=(72/s-1)^2$ is the number of
first-convolution output positions. 
The initiation interval (II) is the HLS-scheduled interval between
successive inference starts.
Cycle counts use the \SI{200}{MHz} inference clock, and utilization is relative
to device totals. Block random access memory (BRAM) usage follows Vitis HLS
BRAM\_18K accounting. DSP, LUT, and FF denote digital signal-processing blocks, lookup tables, and flip-flops.
The uncertainties on $\Rbkg$ are binomial standard errors from 50{,}000
background test events.}
\label{tab:s2dscan}
\tablesetuptight
\begin{adjustbox}{max width=\linewidth}
\begin{tabular}{@{}llccccccccccc@{}}
\toprule
& & & \multicolumn{2}{c}{Latency}
& II
& \multicolumn{4}{c}{Utilization (device)}
& \multicolumn{3}{c}{$\Rbkg$ at $\esig$} \\
\cmidrule(lr){4-5}
\cmidrule(lr){6-6}
\cmidrule(lr){7-10}
\cmidrule(lr){11-13}
Input
& $s$
& $N_1$
& (cycles)
& (\si{\micro\second})
& (cycles)
& BRAM$_{18\mathrm{K}}$
& DSP
& LUT
& FF
& 98\%
& 95\%
& 90\% \\
\midrule

$72\times72\times1$
& 1
& 5{,}041
& 62{,}346
&  311.7
& 62{,}210
& 47.28\%
& 63.05\%
& 61.71\%
& 16.34\%
& \(99.78\pm0.02\)\%
& \(99.90\pm0.01\)\%
& \(99.94\pm0.01\)\% \\

$36\times36\times4$
& 2
& 1{,}225
& 15{,}691
&  78.5
& 15{,}554
& 18.64\%
& 20.81\%
& 25.07\%
& 5.54\%
& \(99.59\pm0.03\)\%
& \(99.80\pm0.02\)\%
& \(99.90\pm0.01\)\% \\

$24\times24\times9$
& 3
& 529
& 8{,}203
&  41.0
& 8{,}066
& 17.82\%
& 16.87\%
& 22.37\%
& 4.89\%
& \(99.11\pm0.04\)\%
& \(99.57\pm0.03\)\%
& \(99.80\pm0.02\)\% \\

$18\times18\times16$
& 4
& 289
& 4{,}352
&  21.8
& 4{,}214
& 19.46\%
& 19.28\%
& 24.47\%
& 4.57\%
& \(98.44\pm0.06\)\%
& \(99.23\pm0.04\)\%
& \(99.61\pm0.03\)\% \\

$12\times12\times36$
& 6
& 121
& 2{,}154
&  10.8
& 2{,}018
& 21.14\%
& 27.31\%
& 29.36\%
& 4.59\%
& \(95.62\pm0.09\)\%
& \(97.97\pm0.06\)\%
& \(98.96\pm0.05\)\% \\

$9\times9\times64$
& 8
& 64
& 1{,}188
&  5.9
& 1{,}055
& 32.62\%
& 40.01\%
& 45.97\%
& 5.83\%
& \(94.45\pm0.10\)\%
& \(97.32\pm0.07\)\%
& \(98.64\pm0.05\)\% \\

\bottomrule
\end{tabular}
\end{adjustbox}
\end{table*}

\subsection{Per-layer precision optimization}
\label{sec:opt:final}

For the selected $s=4$ architecture,
the parameter integer widths were adjusted layer by layer
to reduce quantization error by retaining more fractional bits within the 8-bit format.
The required parameter ranges were determined from the weights and
biases of the selected float32 checkpoint, which also initialized QAT.
The parameter formats and the QAT checkpoint were selected using the
background rejection at $\esig=95\%$ on the validation set.
The hidden-activation format was chosen from the observed non-negative
activation ranges and validation performance.

Accumulator precision was set separately to reduce arithmetic resource use, without retraining the QKeras weights.
The accumulator formats inferred by hls4ml define the uncompressed
reference, which preserves product precision and provides sufficient
integer range for the corresponding sums.
The reduced accumulator widths were selected by requiring exact
output-code agreement with this reference on a fixed 20{,}000-event
validation sample disjoint from the test sample.
Because the accumulators use the default \texttt{ap\_fixed} truncation
and wrap-around modes, a partial sum outside the reduced integer range
wraps rather than saturates.
This occurred in six of the 100{,}000 test
events, whose effect on the scores and trigger decisions is examined
in Sec.~\ref{sec:val:results}.
The final configuration is listed in Table~\ref{tab:hlsconfig}.

\begin{figure}[tbp]
  \centering
  \includegraphics[width=1.0\linewidth]{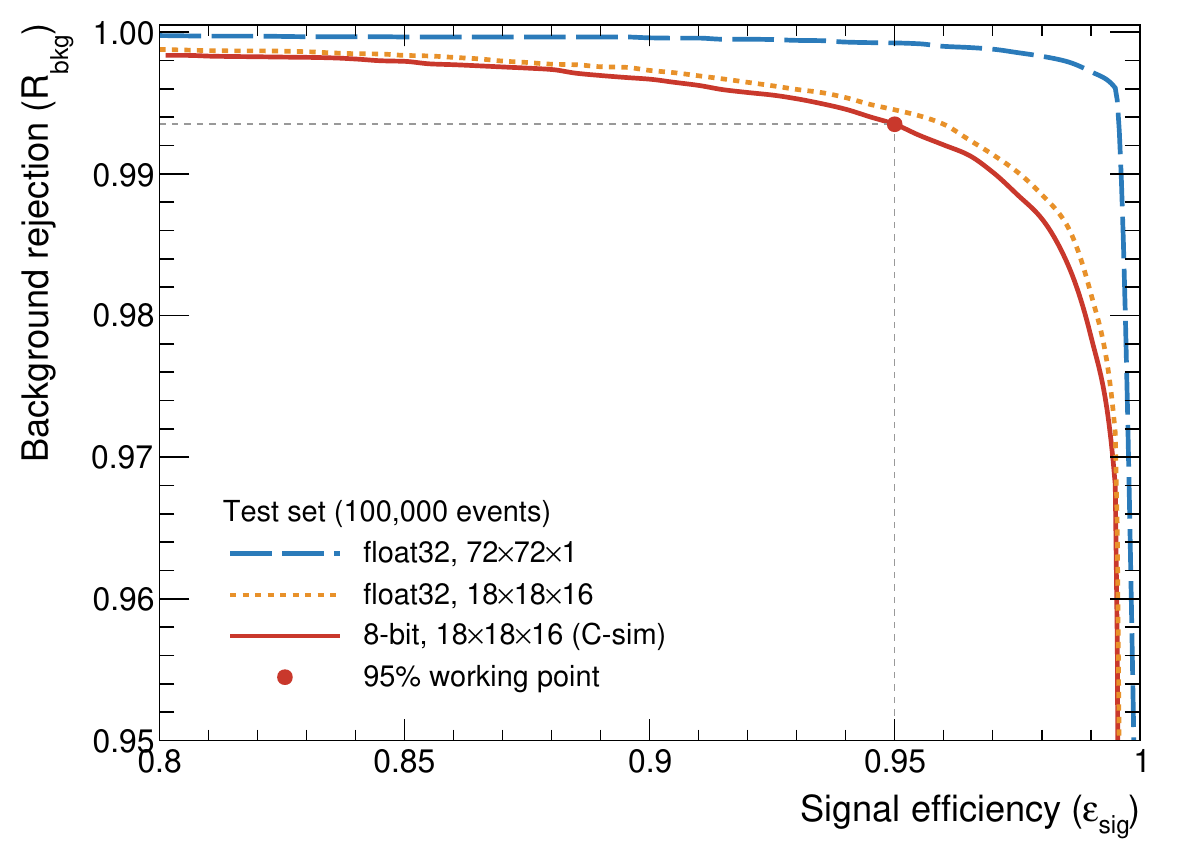}
  \caption{Background rejection versus signal efficiency for
  the float32
  $72\times72\times1$ baseline, the float32 $18\times18\times16$ model, and the
  optimized 8-bit $18\times18\times16$ model, all evaluated on the same
  test sample.
  The fixed-point curve is the Vitis HLS C-simulation result of the implemented model (Sec.~\ref{sec:opt:final}).
  The marker indicates the nominal $\esig=95\%$ working point.}
  \label{fig:roc}
\end{figure}

The operating threshold for the final fixed-point model was selected at
the nominal $\esig=95\%$ point of the 
test-sample receiver operating characteristic (ROC) curve.
Events were accepted if their $\langle16,2\rangle$ output codes were at least
15{,}394, which corresponds to a score threshold of approximately 0.9396.
On the 100{,}000-event test sample, Vitis HLS C simulation of the final
fixed-point model gave a background rejection of $(99.34\pm0.04)\%$
at a nominal signal efficiency of 95\%.
Figure~\ref{fig:roc} compares the final fixed-point curve with the
unfolded and folded float32 baselines.
At $\esig=95\%$, the float32 $18\times18\times16$ model gave
$\Rbkg=(99.45\pm0.03)\%$, providing a same-architecture reference for
the combined effects of quantization and retraining.
For the  final fixed-point model, C synthesis estimated 19.72\%
BRAM$_{18\mathrm{K}}$, 19.25\% DSP, 24.52\% LUT, and 4.44\% FF utilization,
with a maximum latency of 4{,}351 cycles.

\begin{table*}[tbp]
  \centering
  \caption{
  Architecture and numerical configuration of the implemented 
  $18\times18\times16$ CNN.
  All convolutions use unit stride and no padding. Max pooling uses
  $2\times2$ windows with stride two and no padding.
  In $\langle W,I\rangle$, $W$ is the total word length and $I$ the
  integer width, including the sign bit for signed quantities, and the prefix
  $\mathrm{u}$ denotes an unsigned format. 
  The accumulator formats were selected as described in Sec.~\ref{sec:opt:final}.
  The accumulators use the default \texttt{ap\_fixed} truncation and
  wrap-around modes, and the ReLU activations use convergent rounding and
  saturation.}
  \label{tab:hlsconfig}
  \tablesetuptight
  \begin{adjustbox}{max width=\linewidth}
  \begin{tabular}{@{}llrccccc@{}}
    \toprule
    & & & \multicolumn{4}{c}{Precision $\langle W,I\rangle$} & \\
    \cmidrule(lr){4-7}
    Layer
      & Output shape
      & Parameters
      & Kernel
      & Bias
      & Activation/output
      & Accumulator
      & Reuse factor \\
    \midrule

    Input
      & $18\times18\times16$
      & 0
      & ---
      & ---
      & $\langle16,2\rangle$
      & ---
      & --- \\

    Conv0, $2\times2$, 32
      & $17\times17\times32$
      & 2{,}080
      & $\langle8,2\rangle$
      & $\langle8,1\rangle$
      & $\mathrm{u}\langle8,1\rangle$
      & $\langle23,3\rangle$
      & 4 \\

    MaxPool0, $2\times2$
      & $8\times8\times32$
      & 0
      & ---
      & ---
      & $\mathrm{u}\langle8,1\rangle$
      & ---
      & --- \\

    Conv1, $2\times2$, 64
      & $7\times7\times64$
      & 8{,}256
      & $\langle8,1\rangle$
      & $\langle8,1\rangle$
      & $\mathrm{u}\langle8,1\rangle$
      & $\langle16,2\rangle$
      & 16 \\

    MaxPool1, $2\times2$
      & $3\times3\times64$
      & 0
      & ---
      & ---
      & $\mathrm{u}\langle8,1\rangle$
      & ---
      & --- \\

    Conv2, $2\times2$, 128
      & $2\times2\times128$
      & 32{,}896
      & $\langle8,1\rangle$
      & $\langle8,1\rangle$
      & $\mathrm{u}\langle8,1\rangle$
      & $\langle16,2\rangle$
      & 128 \\

    Flatten
      & 512
      & 0
      & ---
      & ---
      & $\mathrm{u}\langle8,1\rangle$
      & ---
      & --- \\

    Dense hidden, 128
      & 128
      & 65{,}664
      & $\langle8,1\rangle$
      & $\langle8,1\rangle$
      & $\mathrm{u}\langle8,1\rangle$
      & $\langle16,2\rangle$
      & 256 \\

    Dense output, 1 + sigmoid
      & 1
      & 129
      & $\langle8,2\rangle$
      & $\langle8,1\rangle$
      & $\langle16,2\rangle$
      & $\langle20,7\rangle$
      & 64 \\

    \midrule
    Total
      & 
      & 109{,}025
      & 
      & 
      & 
      & 
      & \\
    \bottomrule
  \end{tabular}
  \end{adjustbox}
\end{table*}

%==============================================================================
\section{FPGA implementation and hardware validation}
\label{sec:impl}
%==============================================================================

\subsection{Firmware architecture}
\label{sec:impl:firmware}

\begin{figure}[tbp]
  \centering
  \includegraphics[width=1.0\linewidth]{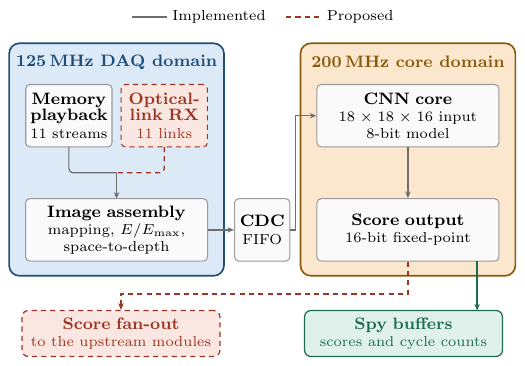}
  \caption{
Block diagram of the DLM firmware described in
Sec.~\ref{sec:impl:firmware}. Solid blocks and paths are implemented and
dashed ones are proposed, including the score-return path to the upstream
modules. FIFO, CDC, and RX denote the first-in, first-out buffer, the
clock-domain crossing, and the optical-link receiver.}
  \label{fig:firmware}
\end{figure}

Figure~\ref{fig:firmware} illustrates the firmware architecture of the
proposed DLM.
Image assembly, including space-to-depth mapping and normalization, operates in the \SI{125}{MHz} DAQ clock domain, whereas
the CNN inference core operates at \SI{200}{MHz}.
A first-in, first-out (FIFO) buffer connects the two stages and
provides the clock-domain crossing (CDC).
In the integrated system, the calibrated energies of the 2,716 CsI
channels would arrive over eleven optical links and be synchronized
by event number.
The firmware maps each energy directly into the folded
$18\times18\times16$ tensor, including the equal subdivision of
outer-crystal energies described in Sec.~\ref{sec:cnn:image}.
The largest cell value, $E_{\max}$, is tracked as the energies are mapped and used
to normalize the completed tensor before it is streamed to the CNN
inference core.
Normalization uses an iterative reciprocal unit with a 16-bit mantissa
and a 5-bit shift, followed by multiplication and rounding to nearest
with ties to even in the $\langle16,2\rangle$ input format.

For each event, the CNN inference core produces a classification
score that is recorded in a spy buffer for offline comparison.
In the integrated system, the score would be returned to the upstream
modules and compared with a programmable threshold to form the trigger decision.
The VPK120 board operates as a standalone module and does
not implement this return path.
Two firmware builds were programmed onto the board: the demonstrator build and the
CNN core build. The demonstrator build replaces the optical-link receivers
with eleven memory playback engines, each emulating the input from one
link. It includes the full processing path from channel energies through
image assembly to CNN inference. This build was used
to measure the latency of individual processing stages and the total
processing latency on the board. The shortest event interval was evaluated
separately in behavioral register-transfer-level (RTL) simulation of this
build. The CNN core build feeds preprocessed $18\times18\times16$ tensors
directly to the network, isolating its arithmetic from the surrounding
input logic. It produced the board inference scores for all 100{,}000
test events used to evaluate the final classification performance.

\subsection{Implementation flow and post-route results}
\label{sec:impl:results}

\begin{figure*}[tbp]
  \centering
  \includegraphics[width=0.95\linewidth]{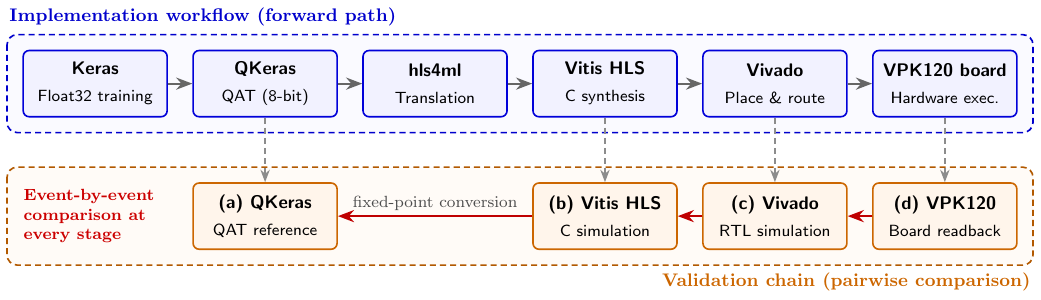}
\caption{
Implementation and validation workflow for converting the trained neural
network into FPGA firmware. The upper row shows the implementation path
from training and quantization-aware training through hls4ml conversion,
C synthesis, place and route, and execution on the VPK120 board. The lower
row shows the event-by-event comparisons among the QKeras reference,
Vitis HLS C simulation, RTL simulation, and board output.}
  \label{fig:workflow}
\end{figure*}

Figure~\ref{fig:workflow} summarizes the implementation and validation flow.
The selected QKeras model of Sec.~\ref{sec:opt} was converted into a
synthesizable C++ description with hls4ml~\cite{hls4ml,hls4ml_cnn}
(version 1.3.0).
The hls4ml implementation uses \texttt{io\_stream}, so feature maps are
transferred as streams and adjacent layers can be pipelined.
Arithmetic parallelism is controlled independently in each learned layer
through its reuse factor (Table~\ref{tab:hlsconfig}), 
chosen to balance latency against multiplier use.
AMD Vitis HLS (version 2023.2)~\cite{vitis} then used this description
to generate RTL code targeting the XCVP1202 device on the VPK120
evaluation board.
The HLS scheduling constraint was \SI{4.5}{ns} with 12.5\% clock uncertainty.
This constraint is tighter than the implemented CNN clock period of
\SI{5}{ns} (\SI{200}{MHz}) to leave scheduling margin for physical
implementation.

The CNN core and demonstrator builds were synthesized, placed, and routed with Vivado
(version 2023.2)~\cite{vivado}.
The routed demonstrator met the \SI{200}{MHz} CNN and \SI{125}{MHz}
DAQ clock constraints, with worst setup slacks of \SI{0.087}{ns} and
\SI{0.926}{ns}, respectively.
Table~\ref{tab:post_route} reports the post-place-and-route resource
utilization for the CNN core and the demonstrator.
DSP engines accounted for approximately 30\% of the device in both builds.
For the CNN core, C synthesis (Sec.~\ref{sec:opt:final}) estimated 19.25\%
DSP and 24.52\% LUT utilization, compared with 29.97\% and 8.85\% after routing.
Only the selected $s=4$ configuration was carried through place and route,
so the estimated resource ranking of the alternative strides has not been verified.

\begin{table}[tbp]
  \centering
  \caption{Post-place-and-route resource utilization of the implemented
8-bit $18\times18\times16$ design on the XCVP1202.
Both builds exclude the optical-link receivers and external
score-return path.
Entries are percentages of the device total, which is given in the
second column. Block RAM is counted in 36-kbit tiles, whereas the
C-synthesis estimates of Table~\ref{tab:s2dscan} are counted in
18-kbit units. CLB denotes configurable logic block.}
\label{tab:post_route}
  \tablesetuptight
  \begin{adjustbox}{max width=\linewidth}
  \begin{tabular}{@{}lccc@{}}
    \toprule
    Resource & Device total & CNN core & Demonstrator \\
    \midrule
    CLB LUTs
      & {900{,}224}
      & 8.85\%
      & {10.15\%} \\
    CLB FFs
      & {1{,}800{,}448}
      & {3.34\%}
      & {4.14\%} \\
    BRAM tiles
      & {1{,}341}
      & {10.74\%}
      & 19.58\% \\
    UltraRAM tiles
      & 677
      & {0.00\%}
      & {0.00\%} \\
    DSP engines
      & {3{,}984}
      & {29.97\%}
      & {30.37\%} \\
    \bottomrule
  \end{tabular}
  \end{adjustbox}
\end{table}

%==============================================================================

\subsection{Numerical and functional validation}
\label{sec:val:results}

Numerical validation compared output scores for the 100{,}000-event test
sample across the QKeras reference, Vitis HLS C simulation, RTL simulation,
and VPK120 board execution. 
All stages received identical preprocessed $18\times18\times16$ tensors 
quantized to $\langle16,2\rangle$.
The QKeras reference uses 8-bit weights and activations with floating-point
accumulation, whereas the other three stages use the fixed-point arithmetic
in Table~\ref{tab:hlsconfig}.
These comparisons distinguish changes introduced by fixed-point conversion
from errors in the firmware implementation.

Scores were compared as integer output codes, with one least significant
bit (LSB) equal to $2^{-14}$ for the $\langle16,2\rangle$ interface.
The test inputs supplied to all four stages
were quantized as $c_x=\mathrm{rint}(x\,2^{14})$, and the QKeras sigmoid
output was placed on the same code grid as
$c_y=\mathrm{rint}(y\,2^{14})$, 
where $x$ is a normalized input-cell energy and $\mathrm{rint}$ denotes
rounding to nearest with ties to even.
Decisions were compared at the operating threshold defined in
Sec.~\ref{sec:opt:final}, which accepts events with $c_y\geq15{,}394$.

C simulation, RTL simulation, and board execution gave identical output
codes for all 100{,}000 test events (Table~\ref{tab:hwval}), so the board
reproduced the final fixed-point classification performance reported in
Sec.~\ref{sec:opt:final} on this sample.
Differences from QKeras did not exceed five output LSB, except for six
events affected by accumulator wrap-around (Sec.~\ref{sec:opt:final}).
These events comprised three signal and three background events, with
a maximum absolute residual of 14{,}937 LSB (0.912 in score).
Restoring the uncompressed accumulators reproduced the QKeras
intermediate values through the sigmoid input for these events.
The remaining nonzero residuals arise from the sigmoid lookup-table
approximation.
At the operating threshold, the fixed-point model rejected one signal
event accepted by QKeras, while no background decisions changed.
This corresponds to a signal-efficiency decrease of 0.002 percentage
points on the test sample.

\begin{table}[tbp]
  \centering
  \caption{
Event-by-event comparison of the QKeras reference, C simulation,
RTL simulation, and VPK120 board execution on the common
100{,}000-event test sample.
Decisions were evaluated using the operating threshold
of output code 15{,}394, inclusive (Sec.~\ref{sec:opt:final}).
Residuals are absolute output-code differences, with one LSB
equal to $2^{-14}$ in score for the $\langle16,2\rangle$ interface.}
  \label{tab:hwval}
  \tablesetuptight
\par\smallskip
  \begin{adjustbox}{max width=\linewidth}
  \begin{tabular}{@{}lcccc@{}}
    \toprule
    & \multicolumn{3}{c}{Event counts} & \\
    \cmidrule(lr){2-4}
    Comparison
      & \shortstack{Decision\\mismatches}
      & \shortstack{Code\\mismatches}
      & \shortstack{Residual\\$>5$ LSB}
      & \shortstack{Max residual\\(LSB)} \\
    \midrule
    QKeras vs.\ C sim.
      & 1
      & 55{,}538
      & 6
      & 14{,}937 \\
    C sim.\ vs.\ RTL sim.
      & 0
      & 0
      & 0
      & 0 \\
    RTL sim.\ vs.\ board
      & 0
      & 0
      & 0
      & 0 \\
    \bottomrule
  \end{tabular}
  \end{adjustbox}%

\end{table}

\subsection{Timing and rate performance}
\label{sec:val:perf}%

Timing is characterized by the CNN inference latency, 
the processing-path latency, and the shortest event interval.
The latency breakdown was measured on the VPK120 board using the
demonstrator build, including its image-assembly and CNN-core stages.
On-chip counters recorded the corresponding cycle counts.
The CNN-core latency from \texttt{ap\_start}
to \texttt{ap\_done} was 3,586 cycles of the \SI{200}{MHz} core clock
(17.9~\si{\micro\second}), 
where \texttt{ap\_start} and \texttt{ap\_done} are the block-level control signals generated by Vitis HLS. 
The processing-path latency was measured on the board from
the first accepted input word of an emulated event to assertion of CNN
\texttt{ap\_done}.
It was 4{,}147 cycles of the \SI{200}{MHz} monitoring clock
(20.7~\si{\micro\second}), including image assembly, the input
clock-domain crossing, and inference. External link transfer and score
return were excluded. Subtracting the inference  latency gives
561 cycles (\SI{2.8}{\micro\second}) for image assembly and its
interface overhead, the contribution quoted in
Table~\ref{tab:latencybudget}.

The event interval was evaluated in behavioral RTL simulation of the
demonstrator using emulated frames reconstructed from preprocessed tensors.
The shortest tested interval at which all events were processed without
loss or reordering was 2{,}260 cycles of the \SI{125}{MHz} DAQ clock,
or \SI{18.1}{\micro\second}, corresponding to
\SI{55.31}{\kilo\hertz}.
This interval was shorter than the \SI{20.7}{\micro\second}
processing-path latency, so successive events were processed with overlap.

The end-to-end estimate in Table~\ref{tab:latencybudget} combines the
measured processing-path latency with estimated external transfers.
It assumes store-and-forward buffering of 128 cycles at each endpoint of both the
input and return paths. The input path requires a further 128 cycles to
transfer the channel energies,
corresponding on average to about two channel energies 
per link and clock cycle, and the return path a further cycle for the score. 
At \SI{125}{MHz}, the assumed input and return contributions are
384 and 257 cycles, corresponding to \SI{3.1}{\micro\second} and
\SI{2.1}{\micro\second}, respectively.
Adding these contributions sequentially gives a projected end-to-end
latency of \SI{25.9}{\micro\second}, leaving a nominal
margin of \SI{6.1}{\micro\second} within the
\SI{32}{\micro\second} R1 budget.
The final model met R2 on the labeled test sample, and the latency
projection supports R1. R3 was assessed for the routed demonstrator,
which excludes the optical-link receivers and external score-return path.

\begin{table}[tbp]
  \centering
  \caption{
Projected end-to-end DLM latency. Entries marked (m) were measured
on the VPK120 board, and entries marked (e) were estimated from the
external-transfer assumptions stated in the text.
The indented rows divide the DLM processing-path latency into its
image-assembly and CNN-inference contributions.}
  \label{tab:latencybudget}
  \tablesetuptight
  \begin{adjustbox}{max width=\linewidth}
  \begin{tabular}{@{}lr@{}}
    \toprule
    Stage & Latency \\
    \midrule
    Link aggregation and transfer to DLM (e)
      & \SI{3.1}{\micro\second} \\
    \addlinespace[2pt]
    DLM processing path (m)
      & \SI{20.7}{\micro\second} \\
    \quad Image assembly (m)
      & \SI{2.8}{\micro\second} \\
    \quad CNN inference (m)
      & \SI{17.9}{\micro\second} \\
    \addlinespace[2pt]
    Score return from the DLM (e)
      & \SI{2.1}{\micro\second} \\
    \midrule
    Projected end-to-end total
      & \SI{25.9}{\micro\second} $<$ \SI{32}{\micro\second} \\
    \bottomrule
  \end{tabular}
  \end{adjustbox}
\end{table}

%==============================================================================
\section{Applications and scalability}
\label{sec:disc}
%==============================================================================

\subsection{Impact on KOTO}
\label{sec:disc:koto}

To estimate the effect on the trigger rate, the final model was 
applied to independent KOTO data collected at a beam power of approximately
\SI{92}{kW}, none of which entered the training, validation, or test samples.
These data contain recorded events from the \kpiee{} trigger line
with no additional offline physics selection.

For the rate study, channel energies were reconstructed from the recorded
CsI waveforms by emulating integration and calibration operations
suitable for online FPGA implementation.
Each waveform was integrated over a 64-sample window after subtracting
the pedestal determined by online tracking in the digitizer.
A four-sample timing window around the nominal pulse timing defined the in-time selection, 
and channels whose pulse peaks fell outside this window were assigned zero energy.
For the remaining channels, a channel-dependent linear calibration
converted the waveform integral to energy, with values below \SI{3}{MeV} also set to zero.
The timing windows and calibration constants were determined from a separate run.
The surviving energies
were mapped and normalized as in Sec.~\ref{sec:cnn:image} and folded as in Sec.~\ref{sec:opt:s2d}, and the
operating threshold defined in Sec.~\ref{sec:opt:final}
was applied unchanged.

The rate sample contains 3{,}865{,}584 events recorded by the \kpiee{} trigger
line with a prescale of ten over 3{,}333 spills, 
corresponding to approximately
1{,}160 recorded events per spill. The operating threshold retained 122{,}743
events, or $(3.175\pm0.009)\%$ of this sample.
This retention was higher than the background retention of $1-\Rbkg=0.66\%$
for the offline-selected test sample enriched in \kpipipi{} decays.
The difference may reflect the broader event composition of the
rate sample, to which this offline selection was not applied.
Correcting for the prescale of ten gives approximately 11{,}600 events
per spill without the DLM and 368 with it, a 31.5-fold reduction
(Table~\ref{tab:physics}). This projection assumes that the measured
retention applies to the unprescaled stream and that the DAQ live time
is unchanged. The DLM would therefore allow the trigger line to operate
without a prescale while recording fewer events per spill than the
current prescaled configuration.

\begin{table}[tbp]
  \centering
  \caption{
Projected event counts and data volumes per spill for the unprescaled
\kpiee{} trigger line at approximately \SI{92}{kW}, with and without
the DLM. The input event count was inferred from independent data
recorded with a prescale of ten, using the same 3{,}333 spills
for both counts. Data volumes assume 597{,}440 bytes per built event
and $1\,\mathrm{GB}=10^{9}$ bytes.}
  \label{tab:physics}
  \tablesetuptight
  \begin{adjustbox}{max width=\linewidth}
  \begin{tabular}{@{}lcc@{}}
    \toprule
    Quantity
      & Without DLM
      & With DLM \\
    \midrule
    Input events per spill
      & $\approx 11{,}600$
      & $\approx 11{,}600$ \\

    Recorded events per spill
      & $\approx 11{,}600$
      & $\approx 368$ \\

    Fraction of input events recorded
      & 100\%
      & 3.18\% \\

    Recorded data volume per spill
      & \SI{6.93}{GB}
      & \SI{0.220}{GB} \\

    Data-volume reduction factor
      & 1
      & $31.5$ \\
    \bottomrule
  \end{tabular}
  \end{adjustbox}
\end{table}

%==============================================================================
\subsection{Extension toward KOTO~II}
\label{sec:disc:koto2}
%==============================================================================

KOTO~II is planned to extend the rare-kaon physics program of KOTO with a
new neutral beam line and a larger calorimeter of higher geometrical
acceptance~\cite{koto2}, which are expected to raise the trigger rate and
increase the need for efficient online selection.
To assess the implications for the DLM implementation, an enlarged-input
benchmark used a $112\times112$ grid matching the KOTO~II calorimeter
dimensions at the \SI{25}{mm} small-crystal pitch.
This grid spans \SI{2.8}{m} and contains approximately 2.4 times as
many cells as the KOTO input.
The simulated signal and recorded background events of
Sec.~\ref{sec:cnn:image} were reused by padding each $72\times72$ input with
20 zero-valued cells on every side.
This permits the hardware requirements and classification performance
to be compared using the same events at the larger input size.
Since the added cells contain no information from a larger detector,
the benchmark does not predict KOTO~II classification performance.

The enlarged-input implementation follows the KOTO architecture, with
space-to-depth folding at $s=7$ mapping the $112\times112$ input to
$16\times16\times49$. The first convolution therefore receives 49 input channels, compared with 16 in the KOTO core.
This larger channel count increases the number of
first-layer parameters and multiplications.
The convolutional and dense widths and the $\langle16,2\rangle$ input and
output interfaces were retained from the KOTO network. Max pooling follows
Conv0 and Conv1 but not Conv2. The resulting model has 113{,}249 trainable
parameters, with the increase over the KOTO model arising entirely from the
additional Conv0 input channels. 
The per-layer parameter formats were derived from the measured ranges
of the enlarged model's float32 checkpoint.
Both the float32 reference and the quantized model were warm-started
from checkpoints with the $16\times16\times49$ architecture, using
the same batch size and regularization as the KOTO models
(Sec.~\ref{sec:cnn:model}).
The hardware mapping uses the same reuse factors and \SI{4.5}{ns} HLS scheduling constraint\ as the KOTO core (Table~\ref{tab:hlsconfig}). The numerical formats also
follow Table~\ref{tab:hlsconfig}, except that the Conv2 bias uses $\langle8,2\rangle$\ and the Conv1, Conv2, and dense-output accumulators use $\langle17,3\rangle$, $\langle18,4\rangle$, and $\langle19,6\rangle$, respectively. The accumulator widths were selected using the output-code agreement
criterion of Sec.~\ref{sec:opt:final}. 
The enlarged core met the \SI{200}{MHz} clock constraint after
place and route, and its board output codes agreed exactly with C simulation for all 100{,}000 test events. 

\begin{table}[tbp]
  \centering
  \caption{KOTO and enlarged-input CNN cores implemented on the XCVP1202.
Rejection was evaluated at a nominal signal efficiency of 95\%
on the common 100{,}000-event test sample, with zero-padding for
the enlarged-input core.
Inference latency was measured on the board at a clock frequency
of \SI{200}{MHz}.
Timing slacks refer to the core builds.
Resources are percentages of the device total.}
  \label{tab:koto2comparison}
  \tablesetuptight
  \begin{adjustbox}{max width=\linewidth}
  \begin{tabular}{@{}lcc@{}}
    \toprule
    & KOTO & {Enlarged input} \\
    \midrule
    \multicolumn{3}{c}{\textit{Model}} \\
    \addlinespace[3pt]
    Input dimensions & $18\times18\times16$ & $16\times16\times49$ \\
    Trainable parameters & 109{,}025 & 113{,}249 \\
    \addlinespace[3pt]
    \multicolumn{3}{c}{\textit{Board measurement}} \\
    \addlinespace[3pt]
    $\Rbkg$ at $\esig=95\%$
      & $99.34\pm0.04$\% & $99.10\pm0.04$\% \\
    Inference latency
      & 3{,}586 cycles & 3{,}471 cycles \\
      &  17.9 \si{\micro\second} &  17.4 \si{\micro\second} \\
    \addlinespace[3pt]
    \multicolumn{3}{c}{\textit{Post-route (Vivado)}} \\
    \addlinespace[3pt]
    Worst setup slack (ns)
      & $+0.134$ & $+0.054$ \\
    Worst hold slack (ns)
      & $+0.010$ & $+0.002$ \\
    CLB LUTs
      & 8.85\% & 13.52\% \\
    CLB FFs
      & 3.34\% & 3.82\% \\
    BRAM tiles
      & 10.74\% & 10.14\% \\
    DSP engines
      & 29.97\% & 46.76\% \\
    \bottomrule
  \end{tabular}
  \end{adjustbox}
\end{table}

The enlarged core used more DSP resources, while its board-measured
inference latency of \SI{17.4}{\micro\second} and block-RAM use
remained close to those of the KOTO core
(Table~\ref{tab:koto2comparison}).
The slightly shorter latency is consistent with fewer first-convolution
output positions (225 compared with 289), while the additional input
channels are processed in parallel using more DSP engines.
The background rejection on the zero-padded KOTO events was lower than
for the KOTO model. This difference reflects combined changes in input
size, folding, per-layer precision, and training settings.
This benchmark demonstrates the hardware feasibility of the tested
enlarged-input configuration, which accommodates about 2.4 times as many
input cells on the same FPGA at a comparable inference latency and with
approximately 47\% DSP utilization.

Evaluating the classification performance and the resulting impact on
the KOTO~II trigger rate requires event samples representative of
its calorimeter cluster distributions and accidental activity under
its detector and beam conditions.

%==============================================================================
\section{Conclusions}
\label{sec:summary}
%==============================================================================

This work demonstrates a real-time FPGA implementation
of a CNN for event classification on full calorimeter images that retain
crystal-level shower information, with KOTO as an experimental benchmark.
The network parameters and hidden activations were quantized to 8 bits
using quantization-aware training, and lossless space-to-depth input
reorganization reduced the estimated inference latency by more than an
order of magnitude.
On the VPK120 board, the final model rejected
$(99.34\pm0.04)\%$ of the recorded background sample
at a nominal signal efficiency of 95\%.
The board output codes agreed exactly with those from the
fixed-point C and RTL simulations for all 100{,}000 test events.
The measured CNN inference latency was \SI{17.9}{\micro\second}.
Including the estimated external transfers, the
projected end-to-end latency of
\SI{25.9}{\micro\second}
lies within the adopted
\SI{32}{\micro\second} DAQ latency budget.

Applying the final model to independent data from the
KOTO \kpiee{} trigger line yielded a projected 31.5-fold reduction in the
trigger rate.
An enlarged-input benchmark with
zero-padded KOTO events shows that a core with about 2.4 times as many
input cells can be implemented on the same FPGA with a comparable
inference latency.

\section*{CRediT authorship contribution statement}

\textbf{J.-J.~Wu}: Methodology, Software, Validation, Investigation.
\textbf{Y.-T.~Lai}: Methodology, Software, Validation, Investigation,
Resources, Supervision, Writing -- review \& editing.
\textbf{Y.-C.~Tung}: Conceptualization, Methodology, Software, Validation, Investigation,
Visualization, Resources, Supervision, Funding acquisition,
Writing -- original draft, Writing -- review \& editing.
\textbf{S.-T.~Chen}: Software, Validation, Investigation, Formal analysis.
\textbf{Y.-S.~Liu}: Methodology, Software, Validation, Investigation,
Formal analysis.

\section*{Declaration of competing interest}
The authors declare that they have no known competing financial interests or
personal relationships that could have appeared to influence the work
reported in this paper.

\section*{Data availability}

Access to the recorded KOTO data is governed by the data policy of the
KOTO Collaboration. The trained models, FPGA configuration and build files,
and validation records used in this study are available from the corresponding
author subject to the applicable access conditions.

\section*{Acknowledgments}

We thank the Collider Electronics Forum (CEF) of the Instrumentation Technology Development Center (ITDC) in the Institute of Particle and Nuclear Studies (IPNS) at KEK for supporting the experimental infrastructure and providing technical assistance. We also thank our colleagues in the KOTO and KOTO~II Collaborations for their support and helpful discussions. This work was supported by the National Science and Technology Council (NSTC), Taiwan, under Grant No. 115-2112-M-017-002.

%==============================================================================
% REFERENCES 
%==============================================================================


\begin{thebibliography}{99}

\bibitem{radovic2018}

A.~Radovic et al., Machine learning at the energy and intensity
frontiers of particle physics, Nature 560 (2018) 41--48,
\href{https://doi.org/10.1038/s41586-018-0361-2}{https://doi.org/10.1038/s41586-018-0361-2}.




\bibitem{belayneh2020}

D.~Belayneh et al., Calorimetry with deep learning: particle
simulation and reconstruction for collider physics,
Eur.\ Phys.\ J.\ C 80 (2020) 688,
\href{https://doi.org/10.1140/epjc/s10052-020-8251-9}{https://doi.org/10.1140/epjc/s10052-020-8251-9}.




\bibitem{csd}

Y.-C.~Tung et al., Suppression of neutron background using deep neural network and Fourier frequency analysis at the KOTO experiment,
Nucl.\ Instrum.\ Methods Phys.\ Res.\ A 1059 (2024) 169010,
\href{https://doi.org/10.1016/j.nima.2023.169010}{https://doi.org/10.1016/j.nima.2023.169010}.



\bibitem{hls4ml}

J.~Duarte et al., Fast inference of deep neural networks in FPGAs for
particle physics, J.\ Instrum.\ 13 (2018) P07027,
\href{https://doi.org/10.1088/1748-0221/13/07/P07027}{https://doi.org/10.1088/1748-0221/13/07/P07027}.



\bibitem{hls4ml_cnn}

T.~Aarrestad et al., Fast convolutional neural networks on FPGAs with
hls4ml, Mach.\ Learn.:\ Sci.\ Technol.\ 2 (2021) 045015,
\href{https://doi.org/10.1088/2632-2153/ac0ea1}{https://doi.org/10.1088/2632-2153/ac0ea1}.



\bibitem{qkeras}

C.N.~Coelho~Jr.\ et al., Automatic heterogeneous quantization of deep neural
networks for low-latency inference on the edge for particle detectors,
Nat.\ Mach.\ Intell.\ 3 (2021) 675--686,
\href{https://doi.org/10.1038/s42256-021-00356-5}{https://doi.org/10.1038/s42256-021-00356-5}.



\bibitem{garnet}

Y.~Iiyama et al., Distance-weighted graph neural networks on FPGAs
for real-time particle reconstruction in high energy physics,
Front.\ Big Data 3 (2021) 598927,
\href{https://doi.org/10.3389/fdata.2020.598927}{https://doi.org/10.3389/fdata.2020.598927}.



\bibitem{belle2_calo_gnn}

I.~Haide et al., Real-time graph neural networks on FPGAs for the
Belle~II electromagnetic calorimeter,
J.\ Instrum.\ 21 (2026) P06039,
\href{https://doi.org/10.1088/1748-0221/21/06/P06039}{https://doi.org/10.1088/1748-0221/21/06/P06039}.




\bibitem{brooke_cnn}

J.~Brooke, E.~Clement, M.~Glowacki, S.~Paramesvaran, J.~Segal,
FPGA implementation of a CNN-based topological trigger for HL-LHC,
Comput.\ Softw.\ Big Sci.\ 9 (2025) 18,
\href{https://doi.org/10.1007/s41781-025-00150-7}{https://doi.org/10.1007/s41781-025-00150-7}.



\bibitem{koto_pnn}

J.K.~Ahn et al.\ (KOTO Collaboration), Search for the
$K_{L}\rightarrow\pi^{0}\nu\bar{\nu}$ decay at the J-PARC KOTO experiment,
Phys.\ Rev.\ Lett.\ 134 (2025) 081802,
\href{https://doi.org/10.1103/PhysRevLett.134.081802}{https://doi.org/10.1103/PhysRevLett.134.081802}.



\bibitem{tung2020}

Y.-C.~Tung, Status and prospect of KOTO-I, J.\ Phys.:\ Conf.\ Ser.\ 1526 (2020) 012026,
\href{https://doi.org/10.1088/1742-6596/1526/1/012026}{https://doi.org/10.1088/1742-6596/1526/1/012026}.



\bibitem{koto2}

J.K.~Ahn et al., Proposal of the KOTO II experiment,
arXiv:2501.14827 [hep-ex] (2025),
\href{https://doi.org/10.48550/arXiv.2501.14827}{https://doi.org/10.48550/arXiv.2501.14827}.



\bibitem{tung2024}

Y.-C.~Tung, Study of $K_L^0\to\pi^0l^+l^-$ at KOTO II,
presentation at the Kaons@J-PARC 2024 workshop, KEK Tokai campus, Japan,
28 July 2024,
\url{https://conference-indico.kek.jp/event/273/contributions/5425/}.



\bibitem{pi0ee_sm1}

G.~Buchalla, G.~D'Ambrosio, G.~Isidori, Extracting short-distance physics
from $K_{L,S}\rightarrow\pi^{0}e^{+}e^{-}$ decays,
Nucl.\ Phys.\ B 672 (2003) 387--408,
\href{https://doi.org/10.1016/j.nuclphysb.2003.09.010}{https://doi.org/10.1016/j.nuclphysb.2003.09.010}.



\bibitem{pi0ee_sm2}

F.~Mescia, C.~Smith, S.~Trine,
$K_{L}\rightarrow\pi^{0}e^{+}e^{-}$ and $K_{L}\rightarrow\pi^{0}\mu^{+}\mu^{-}$: A binary star on the stage of flavor physics,
JHEP 08 (2006) 088,
\href{https://doi.org/10.1088/1126-6708/2006/08/088}{https://doi.org/10.1088/1126-6708/2006/08/088}.



\bibitem{ktev}

A.~Alavi-Harati et al.\ (KTeV Collaboration), Search for the rare decay
$K_{L}\rightarrow\pi^{0}e^{+}e^{-}$,
Phys.\ Rev.\ Lett.\ 93 (2004) 021805,
\href{https://doi.org/10.1103/PhysRevLett.93.021805}{https://doi.org/10.1103/PhysRevLett.93.021805}.


\bibitem{pdg}

F.~Takahashi et al.\ (Particle Data Group), Review of particle physics,
Int.\ J.\ Mod.\ Phys.\ A 41 (2026) 2630011,
\href{https://doi.org/10.1142/S0217751X26300115}{https://doi.org/10.1142/S0217751X26300115}.


\bibitem{csi_calorimeter}

K.~Sato, J.W.~Lee et al., CsI calorimeter for the J-PARC KOTO experiment,
Nucl.\ Instrum.\ Methods Phys.\ Res.\ A 982 (2020) 164527,
\href{https://doi.org/10.1016/j.nima.2020.164527}{https://doi.org/10.1016/j.nima.2020.164527}.

\bibitem{cluster_trigger}

C.~Lin, Y.~B.~Hsiung, Y.~W.~Wah, Y.-C.~Tung, Y.~T.~Luo, M.~Bogdan, The data-acquisition system with cluster-finding trigger at the J-PARC KOTO experiment,
J.\ Phys.:\ Conf.\ Ser.\ 1526 (2020) 012034,
\href{https://doi.org/10.1088/1742-6596/1526/1/012034}{https://doi.org/10.1088/1742-6596/1526/1/012034}.


\bibitem{yolo9000}

J.~Redmon, A.~Farhadi, YOLO9000: better, faster, stronger,
in: Proc.\ IEEE Conf.\ on Computer Vision and Pattern Recognition (CVPR),
Honolulu, HI, 2017, pp.\ 6517--6525,
\href{https://doi.org/10.1109/CVPR.2017.690}{https://doi.org/10.1109/CVPR.2017.690}, arXiv:1612.08242.


\bibitem{spdconv}

R.~Sunkara, T.~Luo, No more strided convolutions or pooling: a new CNN
building block for low-resolution images and small objects,
in: Machine Learning and Knowledge Discovery in Databases. ECML PKDD 2022, Lecture Notes in
Computer Science, vol.\ 13715, Springer, Cham, 2023, pp.\ 443--459,
\href{https://doi.org/10.1007/978-3-031-26409-2_27}{https://doi.org/10.1007/978-3-031-26409-2\_27}, arXiv:2208.03641.


\bibitem{koto_detector}

T.~Yamanaka (KOTO Collaboration), The J-PARC KOTO experiment,
Prog.\ Theor.\ Exp.\ Phys.\ 2012 (2012) 02B006,
\href{https://doi.org/10.1093/ptep/pts057}{https://doi.org/10.1093/ptep/pts057}.


\bibitem{vpk120}
AMD, VPK120 Evaluation Board User Guide, UG1568, version 1.2,
9 April 2025,
\url{https://docs.amd.com/r/en-US/ug1568-vpk120-eval-bd}.


\bibitem{greenlee}

H.B.~Greenlee, Background to $K_L^0\rightarrow\pi^0 ee$ from $K_L^0\rightarrow\gamma\gamma ee$, Phys.\ Rev.\ D 42 (1990) 3724--3731, \href{https://doi.org/10.1103/PhysRevD.42.3724}{https://doi.org/10.1103/PhysRevD.42.3724}.



\bibitem{geant4}

S.~Agostinelli et al., Geant4---a simulation toolkit,
Nucl.\ Instrum.\ Methods Phys.\ Res.\ A 506 (2003) 250--303,
\href{https://doi.org/10.1016/S0168-9002(03)01368-8}{https://doi.org/10.1016/S0168-9002(03)01368-8}.



\bibitem{kl_spectrum}

K.~Shiomi et al., Measurement of $K_L^0$ flux at the J-PARC neutral-kaon beam
line, Nucl.\ Instrum.\ Methods Phys.\ Res.\ A 664 (2012) 264--271,
\href{https://doi.org/10.1016/j.nima.2011.11.010}{https://doi.org/10.1016/j.nima.2011.11.010}.



\bibitem{tensorflow}

TensorFlow Developers, TensorFlow, version 2.14.1, software (2023),
\url{https://github.com/tensorflow/tensorflow/releases/tag/v2.14.1}
(accessed 15 November 2023).



\bibitem{adam}

D.P.~Kingma, J.~Ba, Adam: a method for stochastic optimization,
in: 3rd International Conference on Learning Representations (ICLR),
San Diego, CA, 2015, arXiv:1412.6980,
\href{https://doi.org/10.48550/arXiv.1412.6980}{https://doi.org/10.48550/arXiv.1412.6980}.



\bibitem{vitis}

AMD, Vitis High-Level Synthesis User Guide, UG1399, version 2023.2,
18 December 2023,
\url{https://docs.amd.com/r/2023.2-English/ug1399-vitis-hls} (accessed 23 August 2026).



\bibitem{vivado}

AMD, Vivado Design Suite User Guide: Implementation, UG904,
version 2023.2, 1 November 2023,
\url{https://docs.amd.com/r/2023.2-English/ug904-vivado-implementation} (accessed 23 August 2026).




\end{thebibliography}
\end{document}